\documentclass[%
 reprint,
 longbibliography,
 amsmath,
 amssymb,
 aps,
 pre,
]{revtex4-1}

\usepackage{dcolumn}
\usepackage{bm}

\usepackage{amssymb}
\usepackage{amsmath}
\usepackage{cases}
\usepackage{graphicx,color}
\usepackage{array}
\usepackage{multirow}
\usepackage{makecell}
\usepackage{mathtools}

\definecolor{r}{rgb}{1,0,0}   
\definecolor{g}{rgb}{0,1,0}   
\definecolor{b}{rgb}{0,0,1}
\definecolor{purple}{rgb}{0.808,0.454,0.718}

\begin{document}


\title{Experimentally Testing a Generalized Coarsening Model for Individual Bubbles in Quasi-Two-Dimensional Wet Foams}


\author{A. T. Chieco and D. J. Durian}
\affiliation{Department of Physics and Astronomy, University of Pennsylvania, Philadelphia, PA 19104-6396, USA}


\date{\today}

\begin{abstract}
We present high-precision data for the time evolution of bubble area $A(t)$ and circularity shape parameter $C(t)$ for quasi-2d foams consisting of bubbles squashed between parallel plates.  In order to fully compare with predictions by Roth {\it et al.}\ and Schimming {\it et al.}\ \cite{RothPRE2013,SchimmingPRE2017}, foam wetness is systematically varied by controlling the height of the sample above a liquid reservoir which in turn controls the radius $r$ of the inflation of the Plateau borders.  For very dry foams, where the borders are very small, classic von~Neumann behavior is observed where a bubble's growth rate depends only on its number $n$ of sides.  For wet foams, the inflated borders impede gas exchange and cause deviations from von~Neumann's law that are found to be in accord with the generalized coarsening equation. In particular, the overall growth rate varies linearly with the film height, which decrease as surface Plateau borders inflate.  And, more interestingly, the deviation from $dA/dt\propto (n-6)$ von~Neumann behavior grows in proportion to $nCr/\sqrt{A}$.  This is highlighted definitively by data for six-sided bubbles, which are forbidden to grow or shrink except for the existence of this term.  And it is tested quantitatively by variation of all four relevant quantities: $n$, $C$, $r$, and $A$.
\end{abstract}



\maketitle



\section{Introduction}
Foams are out of equilibrium and made of bubbles that coarsen by the diffusion of gas across the soap films between neighboring bubbles of different pressure \cite{WeaireHutzlerBook}. This occurs in both 2- and 3-dimensional systems, and for any liquid content.  Aging dynamics like coarsening also occur in other cellular systems but the macroscopic nature of bubbles and a known microstructure make foam an ideal system to study \cite{GlazierWeaireRev,StavansRev}. For ideally-dry purely 2-dimensional foams, a bubble's area $A$ changes at a rate that depends only on it's number $n$ of sides according to von~Neumann's law \cite{vonNeumannLaw}:
\begin{equation}
dA / dt = K_o \left( n-6 \right)
\label{vonNeumann}
\end{equation}
where $K_o$ is some rate constant dependent on the physical chemistry of the gas and the surfactant solution. For these idealized 2d dry foams von~Neumann's law is exact because of structural criteria for foams in mechanical equilibrium. In 2d there are three rules called Plateau's laws that govern foam structure and they are as follows: films separating bubbles are circular arcs; films meet in threes at a vertex; the three films at a vertex are separated by equal angles of 120$^o$.  These are true for any foam in a ``dry limit" where it can be considered that the films contain zero liquid and all the liquid phase can be decorated on the vertices \cite{BoltonWeaire1991}.

Bubble coarsening in 2d dry foams has been studied in both experiment  \cite{StavansPRA,StavansPhysicaA, StavansGlazierPRL,deIcazaJAP76} and simulation \cite{GlazieretalPMB,HerdtleArefJFM,SegelEtAlPRE,RutenbergMcCurdyPRE,NeubertSchreckPhysicaA,SeolKimScaling} and general agreement  with Eq.~\ref{vonNeumann} is observed.  Other versions of von~Neumann's law have been developed to describe coarsening behavior of foams embedded in curved surfaces and on flexible membranes \cite{AvronLevine1992,ShimaJPSJ2010,RothPRE2012}. Research on 3d foam coarsening studies bubble volume changes and there exists theory \cite{MullinsActa1989}, simulation \cite{GlazierPRL1993,WangLiuAppPhysLet2008,ThomasColSurf2015}, and experiment \cite{DWS1991,LambertGranerPRL2010, Maret2013}; recently an exact von~Neumann-like expression for individual bubble growth was discovered \cite{MacPhersonSrolovitzNature}.

Foams can age not only by coarsening but also by a combination of drainage and film rupture. The latter can be eliminated by the addition of appropriate surfactants. Stopping drainage is near impossible in experiments unless done in microgravity, but there are ways to study foams with increasing liquid volume fraction. Therefore the coupling of wetness and coarsening is a subject of great interest. In 3d foams an expression for a reduced coarsening rate due to inflated Plateau borders was found in \cite{HutlerWeairePhiloMagLet2000}; similar results are found in subsequent experiments \cite{HilgenfeldtEtAlPRL2001,VeraPRL2002,FeitosaEPJE2008} although the amount of reduced coarsening varies based on the details of the study. For 2d foams the liquid volume fraction is usually considered to exist exclusively at the vertices \cite{BoltonWeaire1991,BoltonWeairePhiloMAgB1992,WeairePhiloMagLet1999,TeixeiraFortesPhilMag2005,ManciniOgueyEPJE2007}. These exactly 2d systems can be studied in theory and in simulation but in experiment the samples are necessarily quasi 2-dimensional. Quasi-2d foams are made of bubbles squashed between two plates so there are both thin films and surface Plateau borders that separate bubbles. The surface Plateau borders run along the top and bottom plates of the cell and can swell with liquid like the Plateau borders in 3d.  The swelling of the edges between bubbles was considered in simulation with a 2d Potts models \cite{FortunaEtAlPRL2012} and in experiments on microfluidic foam \cite{MarchalotEPL2008}. These studies developed empirical formulas to describe the observed coarsening; they do not develop some kind of modified von~Neumann's law whose predictions about the coarsening of individual bubbles rely on the individual bubble-level topology and liquid content of the foam. 

Work in \cite{RothPRE2013} does develop such a modified von~Neumann's law for wet quasi-2d foams. Their new equation is derived by accounting for the size of the Plateau borders due to a higher liquid fraction foam and then going through the same topology based arguments as von~Neumann. The resulting equation is developed for bubbles between two parallel plates separated by a distance $H$ and makes two modifications to Eq.~\ref{vonNeumann}. The first is an overall reduced coarsening rate because gas does not diffuse through the Plateau borders enlarged by the liquid. The second is a violation of von~Neumann's law due to the bubble shape and size. Their model assumes gas does not diffuse at all through the Plateau border but additional work in \cite{SchimmingPRE2017} simulates gas flux through the surface Plateau borders. They find the rate of gas diffusion is not zero and it is set by the geometric mean of the size of the Plateau border and the width of the thin film. Taking the results together from \cite{RothPRE2013} and \cite{SchimmingPRE2017} a new prediction for how the area  $A$ of an $n$-sided bubble changes in time was developed. This generalized coarsening equation is
\begin{equation}
\frac{dA}{dt} = K_o \left(1 - \frac{2r}{H} + \frac{\pi \sqrt{r \ell }}{H}  \right) \left[ \left( n - 6 \right) + \frac{6nCr}{\sqrt{3 \pi A}}  \right]
\label{CoarsenEq}
\end{equation}
where $r=(r_t+r_b)/2$ is the average radius of curvature of the top and bottom surface Plateau borders, $\ell$ is the width of thin films that separate two bubbles, and $C$ is a dimensionless shape parameter ``circularity" of a bubble related to the curvature of the edges of bubble. For an $n$-sided bubble the circularity is 
\begin{equation}
C =\left( \frac{1}{n} \sum_{i=1}^{n}{\frac {1}{\mathcal{R}_i}} \right) \sqrt{\frac{A}{\pi}}
\label{circ}
\end{equation}
where $1/\mathcal{R}_i$ is the curvature of side $i$. Circularity is defined so it is one for circular bubbles, positive for convex bubbles, and negative for concave bubbles. Experimentally, the average and standard deviation of the observed circularities were measured to be approximately  $\left< C \left( n \right) \right> = \left(1 - n/5.73 \right) \pm 0.25$ in the self similar scaling state \cite{RothPRE2013}. Though there is significant variance in circularity between different bubbles of the same $n$, previous work only compared $dA/dt$ data to expectation in terms of the average circularity.

Work from \cite{RothPRE2013} shows that Eq.~\ref{CoarsenEq} accurately predicts the average coarsening behavior of wet foams; the data show there are violations to von~Neumann's law and they are more pronounced for smaller wetter bubbles with $n<6$. One other violation that is apparent from Eq.~\ref{CoarsenEq} and only mentioned in \cite{RothPRE2013} is the fact that the generalized coarsening equation allows for the coarsening of 6-sided bubbles. In this work we show that Eq.~\ref{CoarsenEq} predicts not only average but also individual bubble coarsening behavior with the primary focus being on the latter. From careful reconstructions of individual bubbles we obtain precise measurements of their areas and circularities; the data is used to solve Eq.~\ref{CoarsenEq} and the solutions predict the unique shape dependent coarsening of a bubble with great accuracy.  We present data for individual 6 sided bubbles that coarsen. This behavior is an obvious violation of Eq.~\ref{vonNeumann} and it can only be driven by the bubble shape. Other bubbles with $n \neq 6$ either grow of shrink more slowly than predicted by von~Neumann and in a non linear fashion; this behavior is also predicted by solutions to Eq.~\ref{CoarsenEq} and depends on the bubble size and circularity.


\section{Materials and Methods}
Our experiments consist of making foam in a custom sample cell, allowing it to coarsen until it is quasi 2-dimensional and then taking images of the foam over the course of many hours. The images are then used to reconstruct the foam and from the reconstructions the bubble areas and shapes are determined. We discuss these processes in the following subsections.

\subsection{Experimental Materials}
The foaming solution is 92\% deionized water and 8\% Dawn Ultra Concentrated dish detergent, and has a liquid-vapor surface tension $\gamma=29 \pm 6$ dyn/cm. This solution generates stable foams that do not have any film ruptures. The foam is generated inside a sample cell constructed from two 1.91~cm-thick acrylic plates separated by a spacing $H$=0.32 cm and sealed with two concentric o-rings; additional details about the specifications of the cell are found in \cite{RothPRE2013} and in the supplemental information~\cite{CoarseningSupp}. It features an annular trough that surrounds the the foam and acts a reservoir for excess liquid drained from the foam due to gravity.  The volume of the trough is large compared to the volume of liquid in the foam, so that the height $d$ from the top of the liquid in the reservoir to the middle of the gap between the plates is constant.  The value of $d$ is then set by the amount of liquid sealed into the sample cell, and serves as the key parameter controlling the wetness of the foam.  Specifically, the foam drains into the reservoir, which causes the top and bottom surface Plateau border radii to decrease until capillary and gravitational pressures become equal:
\begin{eqnarray}
	\gamma/r_t &=& \rho g [(d+H/2)-r_t], \label{rtop} \\
	\gamma/r_b &=& \rho g [(d-H/2)+r_b]. \label{rbot} 
\end{eqnarray}
Here $g$ is gravitational acceleration and the terms in square brackets represent the distance from the liquid surface to the respective heights at which the surface Plateau borders begin to flare out from the soap film.  These are the key heights which dominate the border-crossing gas flux \cite{SchimmingPRE2017}.  For a chosen value of $d$, the surface Plateau border radii may thus be computed from these equations for use in Eq.~(\ref{CoarsenEq}).

\begin{figure}[h]
\includegraphics[width=3.2in]{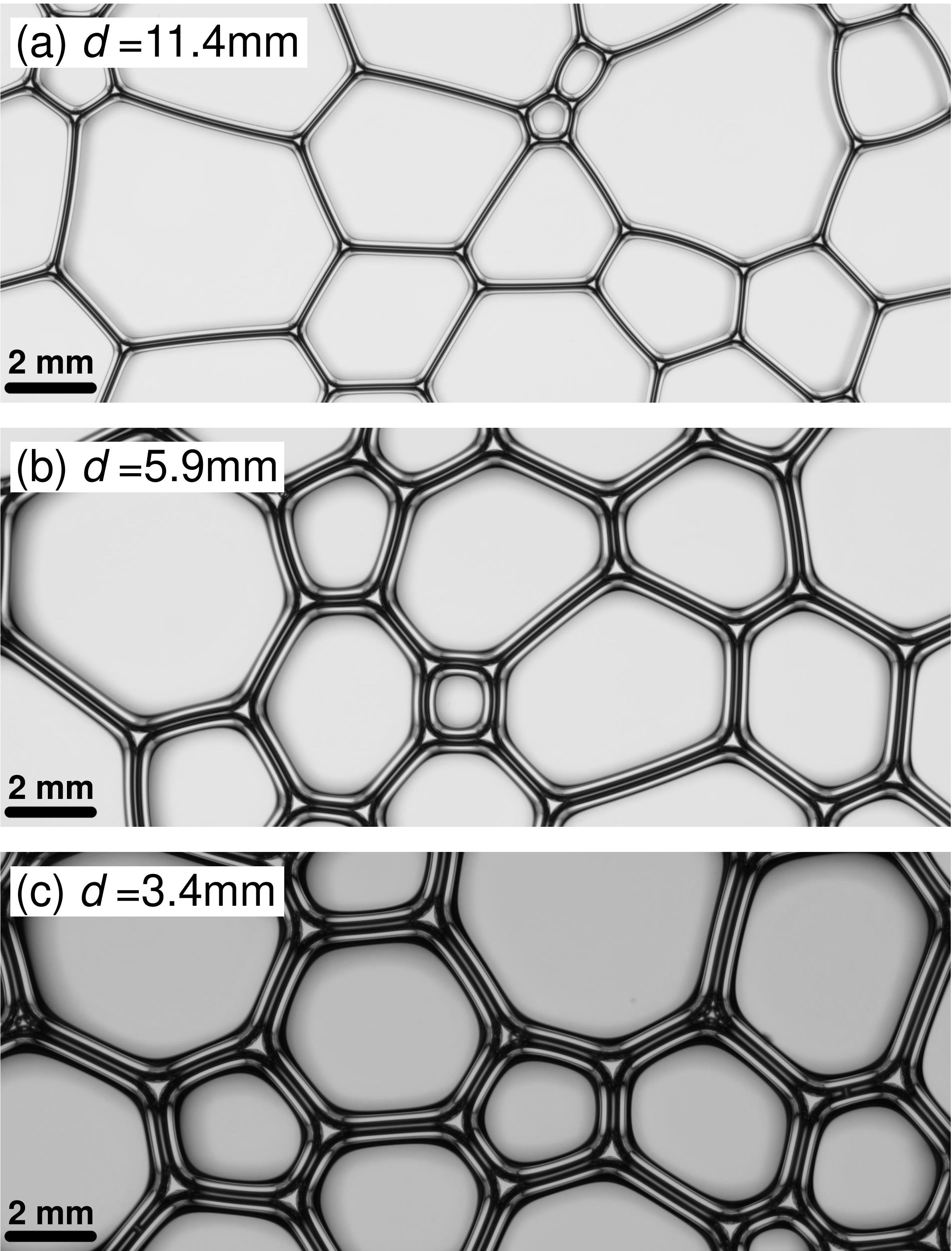}
\caption{Top down view of quasi 2-dimensional foams of various liquid content as indicated by the distance $d$ from top surface of liquid in the sample cell reservoir to the center of the gap between the plates. From (a) to (c) the wetness increases as $d$ gets smaller. The images show the surface Plateau borders along the top plate of the sample cell. The thick surface Plateau borders along the top plate are connected to slightly thicker ones along the bottom plate by thin films. Three surface Plateau borders meet at a surface vertex. The vertices appear bright due to light channeled through vertical Plateau borders that span the gap between both plates.}
\label{Foam_Panels}
\end{figure}

Foams are produced as follows.  First the trough is filled with the desired amount of liquid, then flushed with Nitrogen and sealed. The entire sample cell is vigorously shaken for several minutes until the gas is uniformly dispersed as fine bubbles that are small compared to the gap $H$ between plates.  The foam is thus initially very wet, opaque, and three-dimensional.  Immediately it begins to drain and coarsen, rapidly at first, then progressively more slowly as hydrostatic equilibrium is approached.  After a few hours, the bubbles become large compared to the gap and the coarsening rate is slow compared to drainage.  Thereafter Eqs.~(\ref{rtop}-\ref{rbot}) hold and the foam is quasi-2d as desired for measurement. Fig.~\ref{Foam_Panels} (a-c) shows example images for three such foams with different $d$ and hence different wetness.  There it is evident that the border radii $r$ significantly increase with decreasing $d$.

\subsection{Image Analysis}

\begin{figure}[t]
\includegraphics[width=3.5in]{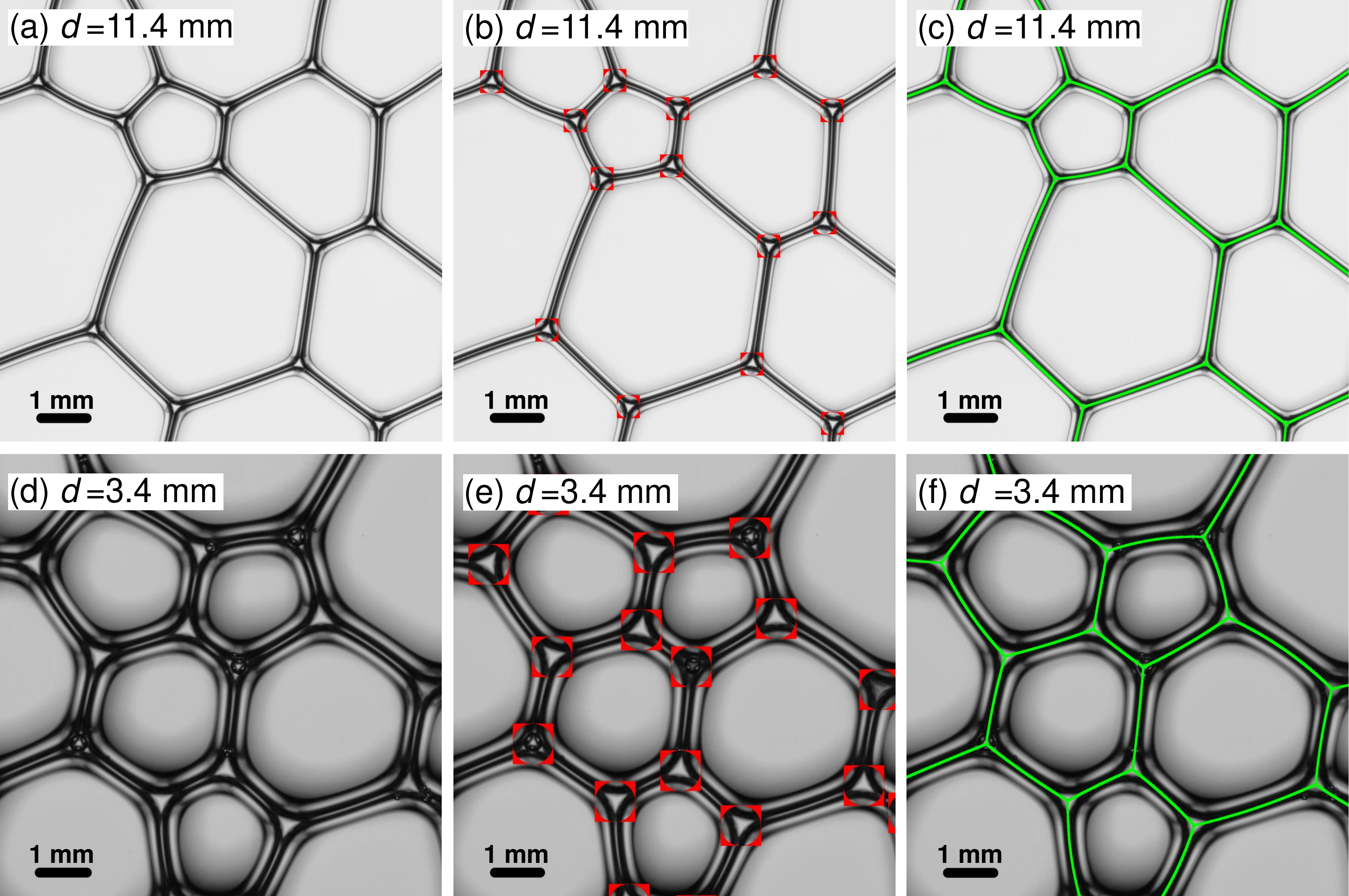}
\caption{Three stages of the foam reconstruction for foams of two different wetnesses as indicated by $d$. Parts (a),(d) show the raw data. Parts (b),(e) show locations and orientations of the vertices determined by a Monte Carlo like method described in the supplement ~\cite{CoarseningSupp}. Parts (c),(f) show the circular arcs that connect pairs of vertices as thick green lines. These arcs reconstruct the film network and they match well with the middle of  corresponding surface Plateau borders. }
\label{FoamProcess}
\end{figure}

After the foam is prepared it is immediately placed 75~cm above a Vista Point A light box and 12.5 cm below a Nikon D90 camera with a Nikkor AF Micro 105mm 1:2.8D zoom lens. The lens is set to full zoom and the entire field of view is 23.3 $\times$ 15.4 mm$^2$. An image is taken every minute for a minimum of 24 hours, but only images of the foam after it coarsens to a quasi-2d state are kept for analysis.  Once the foam enters this state, we must identify the bubbles in the each image to find their shapes and areas. The latter can ostensibly be done by binarizing, skeletonizing and watershedding the pictures of the foam. However our images have features of varying brightness that make the skeletonized images poor representations of the foam and the subsequent watershed basins invalid for measuring the area of bubbles. Instead we have developed an algorithm for reconstructing these wet foams where we find the $(x,y)$ locations of the vertices as well as the orientation of one of the three surface Plateau borders of the vertex with respect to the $x$-axis. For brevity we show our vertex finding method works by displaying in Fig.~\ref{FoamProcess} the found locations of the vertices. The supplemental material thoroughly explains the algorithm for finding the vertices and also includes a movie that demonstrates the process ~\cite{CoarseningSupp}.

The orientations of the surface Plateau borders belonging to a vertex are used to identify the network of neighboring vertices. We know from Plateau's laws that the films in 2d and surface Plateau borders in quasi-2d foams are separated at a vertex by an angle of 120$^o$. Therefore knowing the orientation of one surface Plateau border informs us of the directions of the others. We know precisely where to investigate in order to find the three neighbors of a vertex. After the neighbors are determined for each vertex,  we connect them to recreate the film network of the foam. 

Another of Plateau's laws is that films connecting the vertices in quasi 2-dimensional foam are arcs of circles. The center and radius $(x_c, y_c, \mathcal{R})$ of the circles that connect any pair of vertices are defined by the vertex locations and a point midway between the two vertices in the middle of the surface Plateau border. The method used to determine this third point is presented in the supplemental material ~\cite{CoarseningSupp}. Additionally the reconstructed film network is adjusted to better satisfy Plateau laws; explanations of this process and a movie representing the evolution of the reconstructions are also included in the supplemental material ~\cite{CoarseningSupp}. In Fig.~\ref{FoamProcess} (c) and (f) we show the circular arcs that reconstruct the film network.

With the film network carefully reconstructed we can finally determine the areas and shapes of the coarsening bubbles. We first identify which vertices belong to a bubble. The bubble area is then calculated in two steps using first the location of the vertices and then using the equations of the circular arcs that connect them. The bubble is initially treated like a polygon where the vertices are connected by straight lines. This treatment gives a polygonal area of $\alpha=\sum{\left(x_i y_{i+1}+x_{i+1} y_i\right)}/2$ where the sums are between all pairs of connected vertices belonging to a bubble. Because the vertices are actually attached by arcs of circles and not straight lines we then account for the  area under the circular arcs; the bubble area is in total its polygonal area plus or minus the area under each of the $n$ circular arcs of the bubble if the arc bends away or towards the centroid of the bubble, respectively. Once the bubble area is known we use it along with the values of $\mathcal{R}$ for each of the $n$ sides of the bubble to evaluate Eq.~\ref{circ}. This is done for all bubbles in an image and for all images.

Finally we measure the uncertainty in the areas and circularities. These uncertainties account for how well the foam is reconstructed and to determine them we re-fit the films connecting neighboring vertices. The new fits are done on three points: two of the points are the vertex locations slightly shifted so the distance between them increases; the third remains in the bright band in the middle of the film but is shifted to maximize the distance from each vertex. This provides new values for the center and radius $(x_c', y_c', \mathcal{R'})$ and these are used to find new areas $A'$ and circularities $C'$. The uncertainties are then $\Delta A = A-A'$ and $\Delta C = C - C'$ where $A$ and $C$ are the originally calculated values of the area and circularity of each bubble. Once the areas, circularities, and uncertainties for each are calculated, all bubbles are tracked using standard particle tracking procedures.

\section{Coarsening Rates}

\begin{figure}[t]
\includegraphics[width=3in]{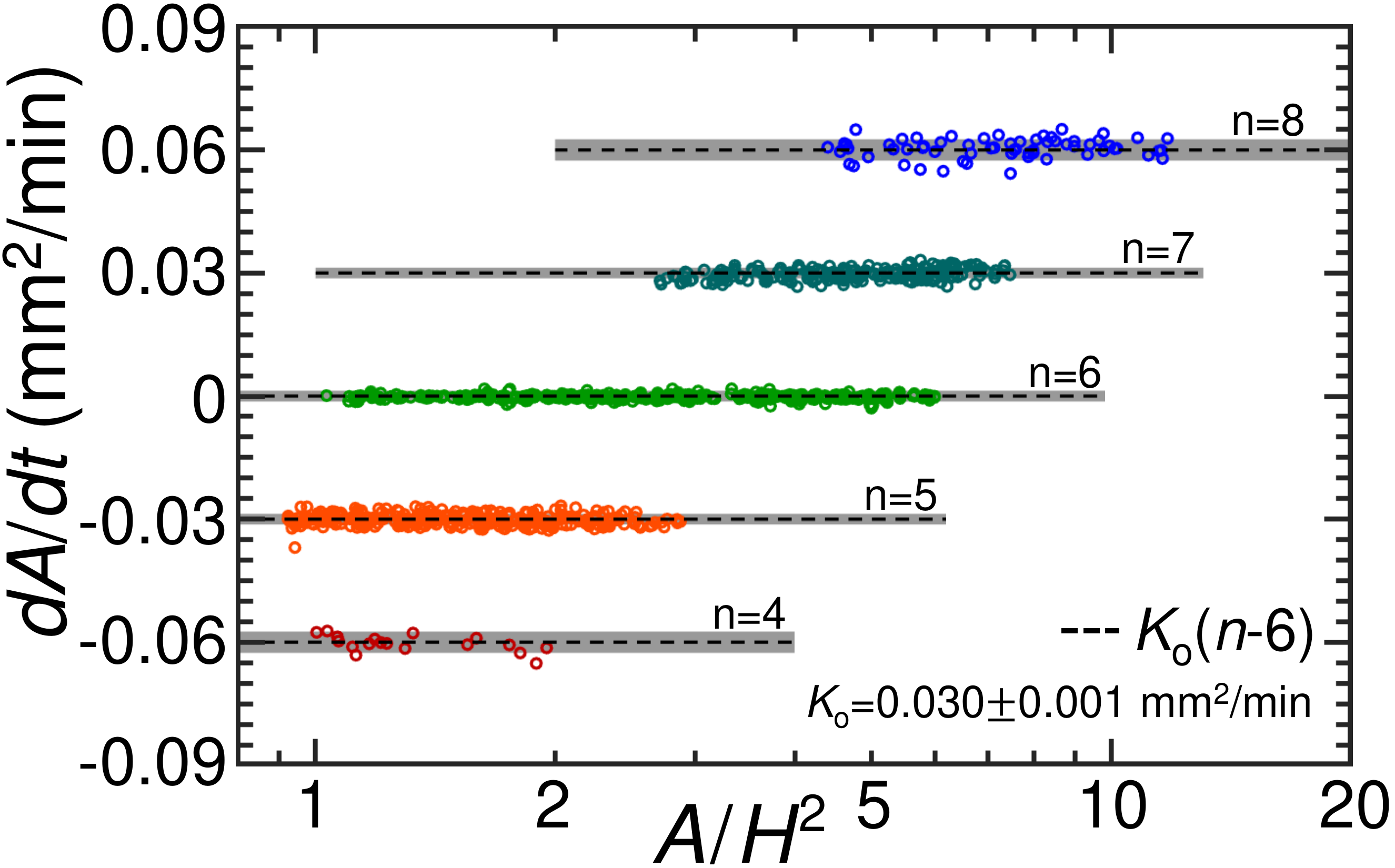}
\caption{Rate of change of bubble area $A$ versus $A$ divided by the square of the gap $H$ between plates, for a dry foam consisting of bubbles with different numbers of sides as labeled by $n$. The von~Neumann expectation $dA/dt=K_o(n-6)$ is plotted for $K_o=0.030 \pm 0.001$ mm$^2$/min, as shown by horizontal dashed lines and gray swaths.}
\label{Dry_Coarsening}
\end{figure}

\begin{figure}[t]
\includegraphics[width=3in]{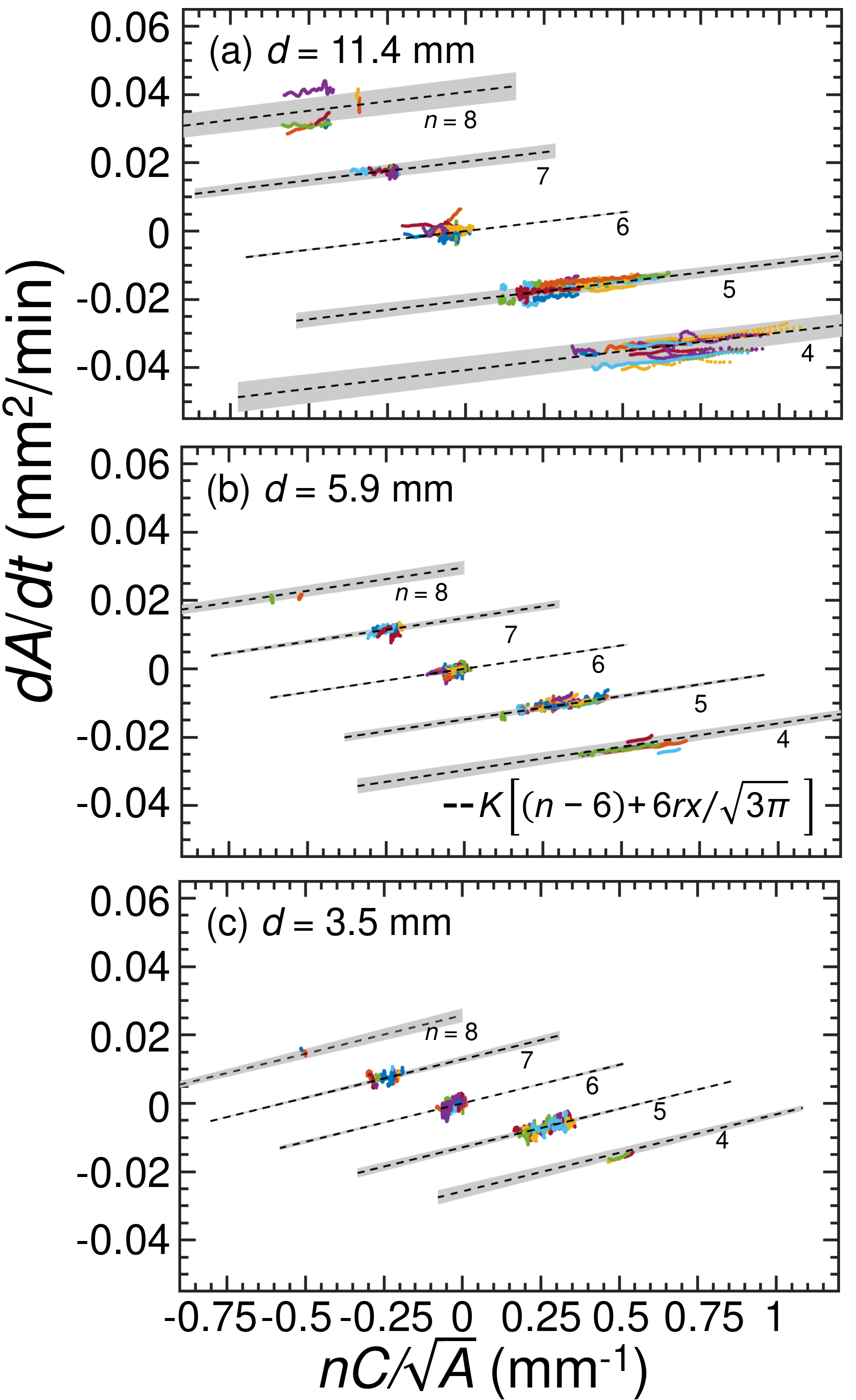}
\caption{Coarsening rate versus number of sides multiplied by the circularity and divided by the square root of the area of individual bubbles; data are shown for foams with various liquid content as labeled by $d$. Individual bubble coarsening data are shown as colored dots and the $x$-axis values are calculated using the bubble specific values of $C \left( t \right) / \sqrt{A \left(t \right)}$. The black dashed lines show the expectation for the generalized coarsening equation; they are evaluated by making simultaneous fits to the coarsening rates for all bubbles in each wetness where the reduced coarsening rate $K$ and the average radius of curvature of the Plateau borders $r$ are fit parameters. The gray swaths show the equation evaluated using $K \pm \Delta K$ and $r \pm \Delta r$. }
\label{Wet_Coarsening}
\end{figure}

Having tracked individual bubbles, we observe how their areas change throughout their lifetime. Recall that von~Neumann's law says for 2 dimensional foams with no wetness that the coarsening rate of a bubble should depend only on its number of sides; bubbles with $n>6$ grow, bubbles with $n<6$ shrink and bubble with $n=6$ do not have their area change. Eq.~\ref{CoarsenEq} generalizes coarsening behavior for quasi-2d wet foams where the wetness of the foam along with the size and shape of a bubble will have an effect. Coarsening rates for bubbles in quasi-2d foams are shown in Fig.~\ref{Dry_Coarsening} for foams in a very dry limit and in Fig.~\ref{Wet_Coarsening} for wet foams of varying liquid content.

To test for von~Neumann like behavior we make very dry foams by standing the sample cell so that the plane of the foam is vertical. Drainage results in all the liquid pooling to the bottom of the cell and bubbles far from the liquid do not have any enlarged Plateau borders. A very conservative estimate for this distance is 1 cm above the bath which is about 4$\times$ the capillary length and only bubbles this distance and above the liquid pool are analyzed. For dry foams it is easier to acquire information about the bubble areas than for the wet foams. The areas of dry bubbles are determined from a process where we binarize, skeletonize, and watershed images of the foam. Bubbles are the watershedding basins of the skeletonized images and the number of pixels within each basin is converted into the bubble area. 

Individual bubble tracks show areas that change linearly with time and to find the coarsening rate for each bubble we fit lines to the data. The values of $dA/dt$ are plotted versus bubble area in Fig.~\ref{Dry_Coarsening}. It is evident that the coarsening rates are the same for all bubbles with the same number of sides and the coarsening rate for $n$-sided bubbles follows von~Neumann's law $dA/dt=K_o \left(n-6\right)$. The rate constant $K_o$ is found to be $K_o = 0.030 \pm 0.001$ mm$^2$/min. 

Turning to our main interest in wet foams, we find by contrast with the dry case that bubbles with the same number of sides do not all coarsen at the same rate. This is shown in Fig.~\ref{Wet_Coarsening} which plots $dA/dt$ not versus area but rather versus the quantity $x=nC/\sqrt{A}$ that controls the deviation from von~Neumann behavior in Eq.~(\ref{CoarsenEq}).  For this, the values for the $x$-axis are determined from the raw values of $C\left(t\right)$ and $A\left(t\right)$ taken for individual bubble tracks, and also the number $n$ of sides found from the reconstruction of the network.  The coarsening rates on the $y$-axis are the  numerical derivatives of the area data smoothed over a Gaussian window. Indeed, we find in Fig.~\ref{Wet_Coarsening} that $dA/dt$ is not constant for a given $n$ but rather varies linearly in $x$ as predicted.  Also as predicted, the slope varies with wetness independent of $n$.  Thus there is good qualitative agreement with expectation, which may now be tested more quantitatively:

According to Eq.~(\ref{CoarsenEq}), the separation and slope of the data clusters are set respectively by the values of $K=K_o\left( 1- 2r/H + \pi \sqrt{ r\ell } / H \right)$ and the average radius of curvature $r$ of the surface Plateau border. To find $K$ and $r$, these parameters are adjusted to simultaneous fit the coarsening rates for all the bubbles in each wetness.  Excellent fits are achieved, as illustrated by the dashed lines in Fig.~\ref{Wet_Coarsening}.  The gray swaths show the fitting equation evaluated across the the acceptable range of fitting parameters, $K\pm\Delta K$ and $r\pm\Delta r$.

\begin{figure}[t]
\includegraphics[width=3in]{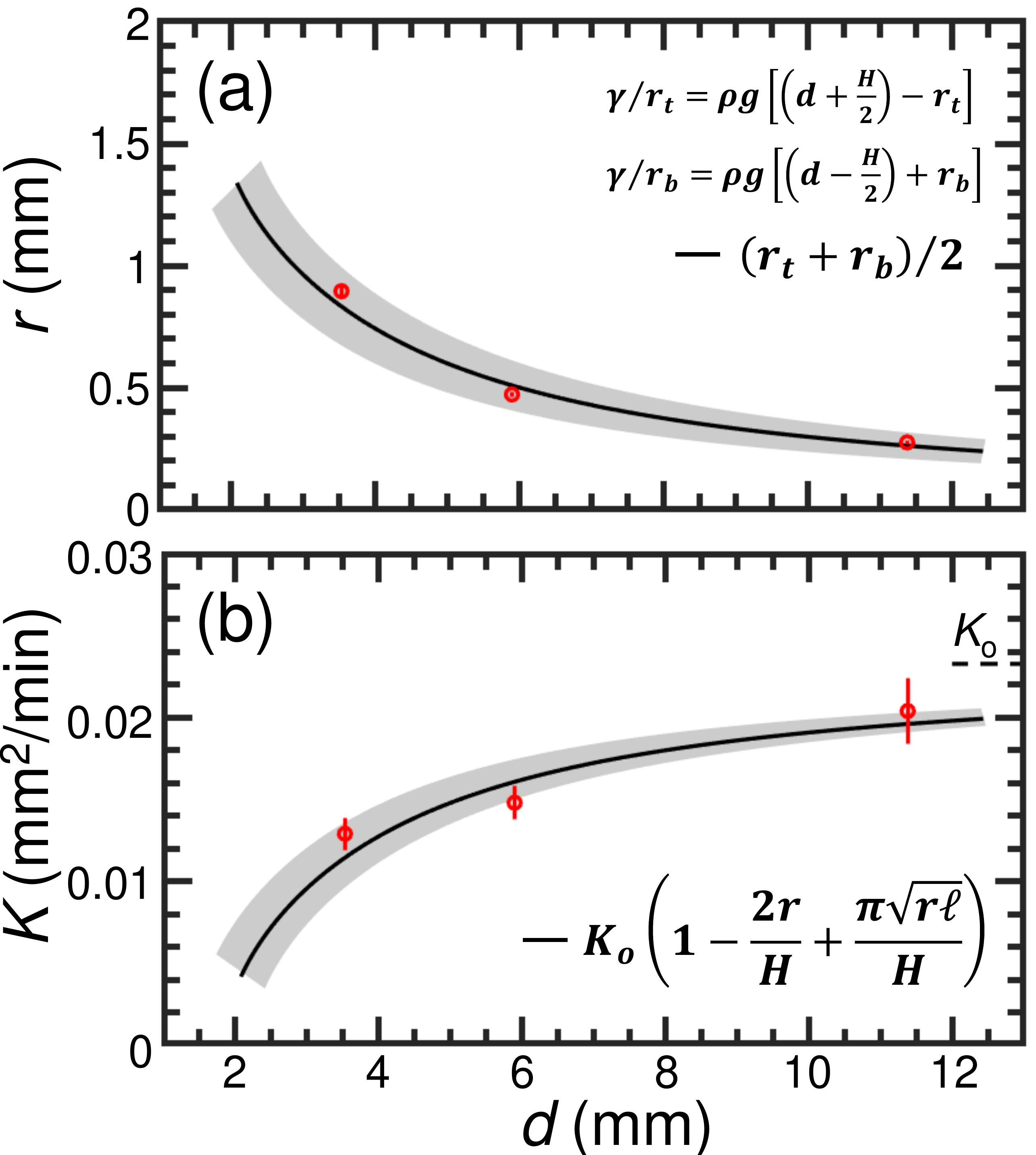}
\caption{The average Plateau border radius (a) and reduced coarsening rate (b) versus the distance $d$ from top surface of liquid in the sample cell reservoir to the center of the gap between the plates.  Data are shown as points, and expectations are shown as solid curves with a gray swath reflecting uncertainty in $\gamma$, $\rho$, $d$ and $\ell$. Part (b) plots the value of $K_o$ used to evaluate the expectation as a dashed line. Note that as $d$ increases, the foam becomes drier and hence $r$ decreases and $K$ increases in good accord with expectation.}
\label{r_and_K_vals}
\end{figure}

To complete the analysis, the fitting parameter results are plotted versus $d$, which controls wetness, and compared with expectation in Fig.~\ref{r_and_K_vals}. In part (a), results for the average Plateau border radius $r$ decrease with increasing $d$ for drier foams.  The expectation for $r$ is shown as a solid curve surrounded by a gray swath that represents the uncertainties in $\gamma$, $\rho$, and $d$.  There are no fitting parameters, and even so the agreement is essentially perfect.  In part (b), results for $K$ increase with increasing $d$ for drier foams and the expectation is similarly shown.  Now the overall rate $K_o$ is adjusted to give a good fit to the data.  This yields $K_o=0.023\pm 0.002$~mm$^2$/min, taking the film thickness from across a reasonable wide range of values, $10^{-5}{\rm \ mm} < \ell < 10^{-3}$~mm \cite{BergeronRadke1992,KralchevskiNikolov1990,ZhangSharma2018}.  The result is somewhat smaller than the value $K_o=0.030\pm0.001$~mm$^2$/min measured in Fig.~\ref{Dry_Coarsening} for a perfectly dry foam using the usual von~Neumann equation.  The source of discrepancy is not known, but could arise by a slight change in the physical chemistry of the solution.  Nevertheless, the parameters $r$ and $K$ give excellent fits in Fig.~\ref{Wet_Coarsening} to the expected variation with wetness, further demonstrating the validity of the generalized coarsening equation.

\section{Individual Bubble Coarsening}

In this final section we highlight that bubble shape drives the violation of von~Neumman's law, by returning to the raw data for bubble area and circularity versus time for a few individual bubbles in the above analyses.  We begin with six-sided bubbles, some of which grow and some of which shrink as seen by careful inspection of the sign of $dA/dt$ data in Fig.~\ref{Wet_Coarsening}.  This effect, and its analogue for $n\ne6$, is more obvious and dramatic in $A(t)$ versus $t$ data for individual bubbles as follows.

\subsection{Bubbles with $n= 6$}

\begin{figure*}[ht]
\includegraphics[width=7in]{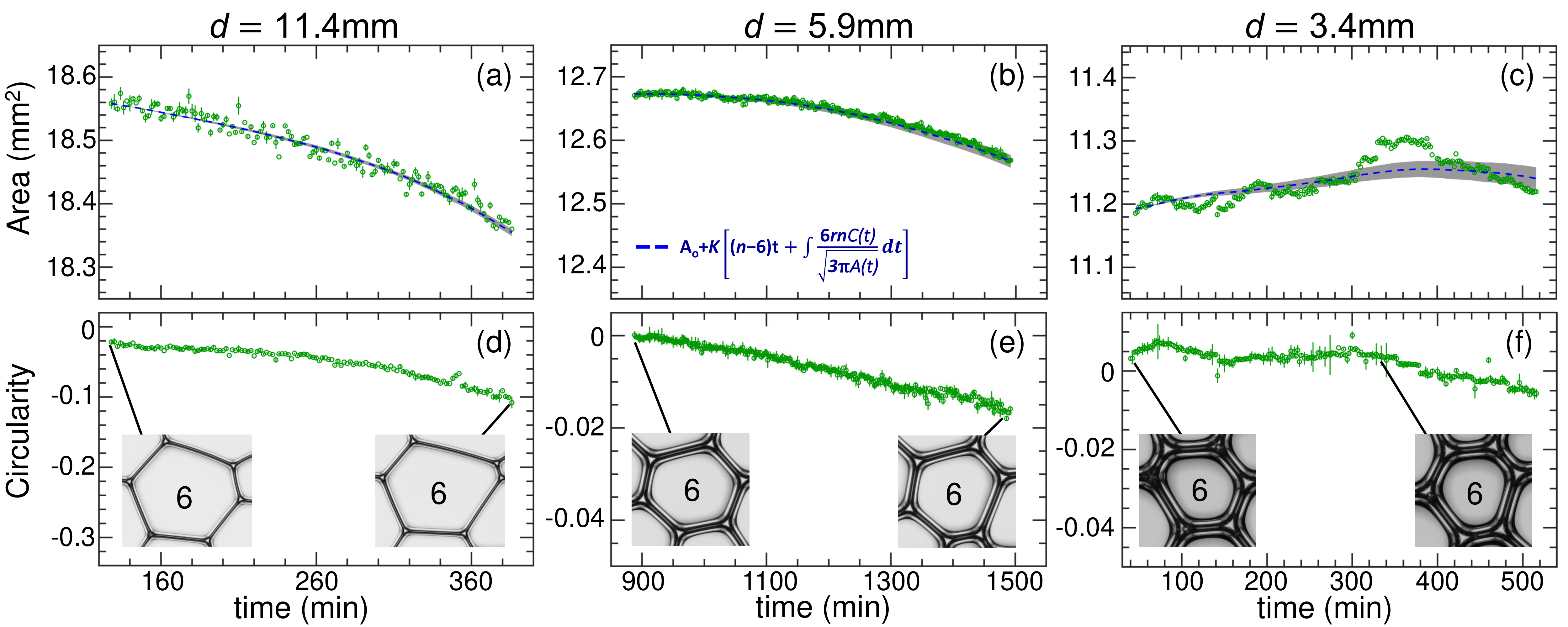}
\caption{The area (a-c) and circularity (d-f) versus time for individual 6-sided bubbles. The bubbles come from foams of increasing wetness from left to right as labeled by decreasing $d$. The images in parts (d-f) show the bubble at the times pointed to by the black lines. The von~Neumann expectation is that 6-sided bubbles do not change their area in time but part (a)/(b) shows 6-sided bubbles that shrink and (c) shows a bubble that grows. These area changes are driven by the bubble circularity which is negative at all times in parts (a)/(b) and positive during the bubble growth in part (c). The blue dashed lines are the numerically integrated solutions to the generalized coarsening equation using the bubble circularity data as well as values of $K$ and $r$ corresponding to the bubble wetness. The gray swaths are generated similarly, incorporating statistical uncertainty in $C$ as well as uncertainty in the values of $K$ and $r$. }
\label{Six_Sided_Coarsening}
\end{figure*}

According to von~Neumann's law, the area of 6-sided bubbles in a dry foam should not change in time. By contrast, for wet foams, the expectation of Eq.~\ref{CoarsenEq} for 6-sided bubbles reduces to $dA/dt = 6KnrC\left(t\right) / \sqrt{3 \pi A\left(t\right)}$.  Therefore, the magnitude and sign of the circularity shape parameter $C\left( t \right) / \sqrt{A\left( t \right)}$ determines whether a 6-sided bubble grows or shrinks and at what rate. Data for the area of examples bubbles in foams of various wetness are shown in Fig.~\ref{Six_Sided_Coarsening}, where the area is seen to either decrease as in parts (a)/(b) or to grow as in part (c). The corresponding circularities are plotted in Fig.~\ref{Six_Sided_Coarsening} (d-f).  There we see $C \left( t \right)<0$ in (d)/(e) for bubbles that shrink and $C \left( t \right)>0$ in (f) for the bubble that grows.  This shows there is good qualitative agreement with Eq.~\ref{CoarsenEq} and the coarsening behavior. Additional data in the supplement for $A\left(t\right)$  and $C\left(t\right)$ versus time  ~\cite{CoarseningSupp} show more examples of qualitative agreement between the change in area and bubble shape for many 6-sided bubbles from foam with $d=5.9$mm.

To demonstrate the quantitative validity of Eq.~(\ref{CoarsenEq}) for $dA/dt$ for these three six-sided bubbles, we numerically integrate it using the displayed $C(t)$ data in order to obtain $A(t)$ versus $t$ along with the fitted values of $K$ and $r$ discussed above.  The resulting predictions for $A(t)$ versus $t$ are displayed as dashed curves with a surrounding gray swath that reflects statistical uncertainties in $K$, $r$, and $C$.  Evidently, the agreement is remarkably good.  Of course this is expected based on the success of the fits in the previous figures. Nevertheless it is a powerful demonstration that von~Neumann's law is indeed violated for wet foams according to prediction in terms of the bubble shape.  Note that the agreement is accurate at the level of $\approx0.01$~mm, and that the comparison was made possible by the high precision of our data.

These results raise a new question: What controls the value and time-evolution of a bubble's circularity shape parameter and hence whether $A$ grows or shrinks?  To begin exploring this issue, we examine the photographs of the example bubbles shown in Fig.~\ref{Six_Sided_Coarsening} at early and late times.  In these it is not possible to visually discern the area changes. But, in (d), it is nevertheless apparent that the shortest side becomes much shorter and more highly negatively curved. Since the Eq.~(\ref{circ}) definition of $C$ features an unweighted sum of curvature for each side, the very-short very-curved film contributes very strongly to $C$ and hence is responsible for it being both negative and a decreasing function of time.  More generally, six-sided bubbles often have a small few-sided bubble as neighbor that shares a short film that shrinks and becomes more curved with time.  Thus we find $C(t)$ tends to be negative and decreasing for many six-sided bubbles.  Six-sided bubbles with $C>0$, that grow with time as in part (c), exist but are more rare. We leave it to future studies to further consider the distribution and evolution of shape parameters in terms of nearest-neighbor size and shape correlations.

\subsection{Bubbles with $n \neq 6$}

While the way bubble shape drives violation of the von~Neumann law is most evident for six-sided bubbles, it can also be seen for bubbles with other side numbers $n$. According to Eq.~(\ref{CoarsenEq}) the coarsening rates for bubbles with the the same $n$ are different from one another depending on the individual bubble shape; how the circularity of a bubble affects its coarsening is demonstrated qualitatively in the supplement with plots showing $A\left(t\right)$ and $C\left(t\right)$ versus time for many $n$-sided bubbles from foam with $d=5.9$mm ~\cite{CoarseningSupp}. 

\begin{figure}[ht]
\includegraphics[width=3.4in]{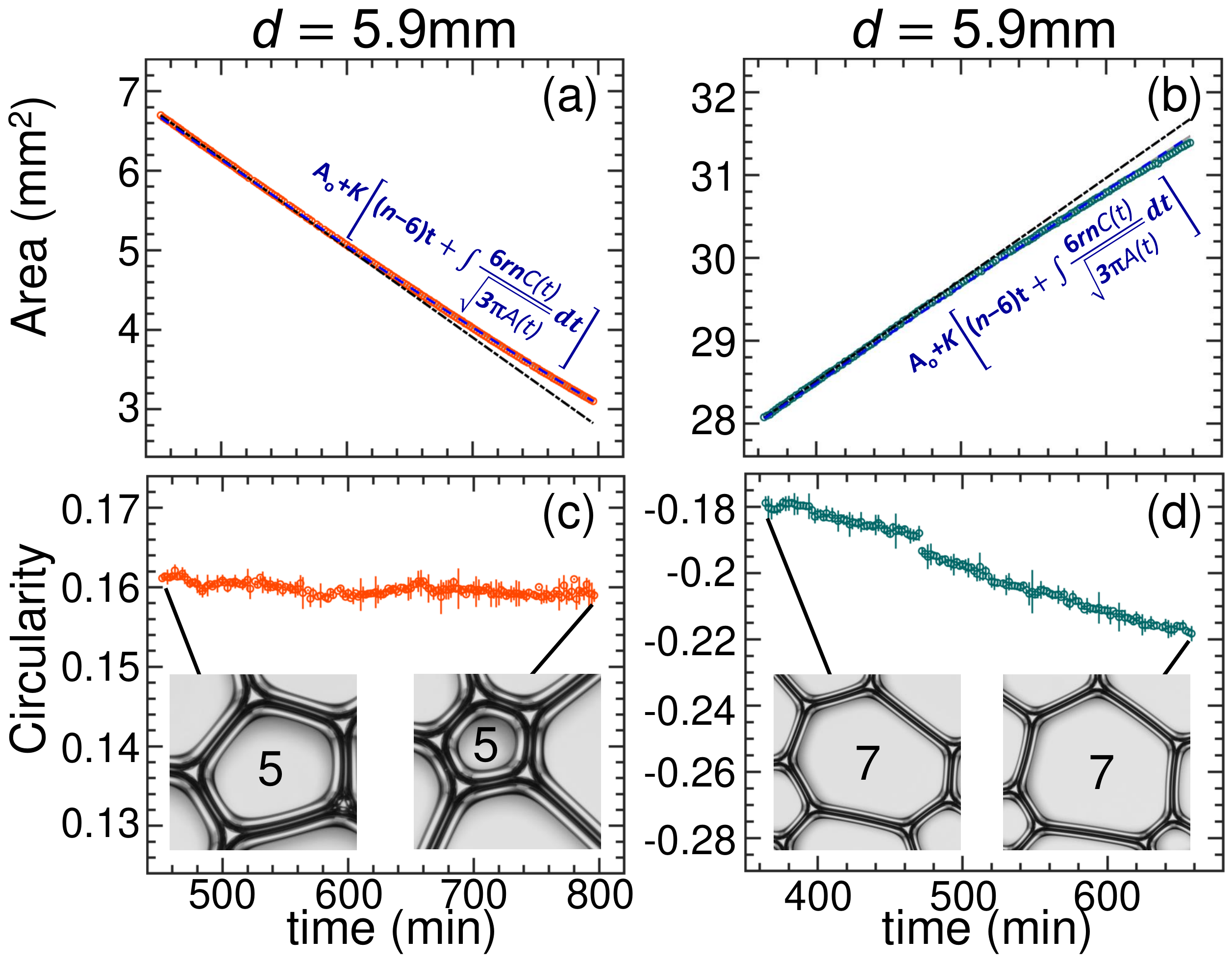}
\caption{The area (a)-(b) and circularity (c)-(d) versus time for individual bubbles with $n$ number of sides as labeled in the bubble images; both bubbles come from a foam with the same wetness indicated by $d$. The images in (c) and (d) show the bubble at the times pointed to by the black lines. In parts (a) and (b) the black dash-dotted lines are tangent to the data at early times and demonstrate the the area changes for these bubbles is non-linear. This behavior is due to the bubble circularity where 5-sided or 7-sided bubbles shrink or grow more slowly because $C\left(t\right)>0$ or $C\left(t\right)<0$, respectively. The blue dashed lines are the numerically integrated solutions to the generalized coarsening equation using the bubble circularity data as well as values of $K$ and $r$ corresponding to the bubble wetness. The gray swaths are generated similarly, incorporating statistical uncertainty in $C$ as well as uncertainty in the values of $K$ and $r$.}
\label{Five_Seven_Sided_Coarsening}
\end{figure}

Quantitatively this is demonstrated for one five-sided and one seven-sided bubble in Fig.~\ref{Five_Seven_Sided_Coarsening} and for other bubbles in ~\cite{CoarseningSupp}. Just as in Fig.~\ref{Six_Sided_Coarsening}, the top and bottom rows respectively show area and circularity data versus time, along with photographs of the bubbles at early and late times.  For $n=5$ and for $n=7$ the bubbles respectively shrink and grow, nearly linearly with time as expected from the von~Neumann law.  Indeed the area change is evident in the photographs.  However, in both cases the area change is slightly slower than linear. And in fact this deviation is perfectly captured by numerical integration of the coarsening equation using the $C(t)$ data, exactly as done for the six-sided examples.  Now the sign of the deviation is more clear:  Five-sided bubbles always have positive circularity, while $(n-5)$ is negative; therefore they shrink more slowly than von~Neumann.  Similarly, seven-sided bubbles always have negative circularity, while $(n-6)$ is positive; therefore they grow more slowly than von~Neumann.  For the example $n=5$ bubble, the circularity is roughly constant as it shrank.  By contrast, the circularity of $n=7$ example bubble decreased as two of its shorter sides became even shorter and more curved -- similar what was seen for typical $n=6$ bubbles. 

\section{Conclusion}

In this work we show that the generalized coarsening equation describes both the average coarsening behavior of bubbles in a wet foam and predicts changes to the area of individual bubbles. To show the average behavior follows Eq.~\ref{CoarsenEq} we take a simultaneous fit to the data of different $n$-sided bubbles where the fit parameters are the reduced coarsening rate $K$ and the radius of curvature of the Plateau borders $r$. The $r$ values from the fit agree with the calculated values of $r$ values determined from Eq.~\ref{rbot} and Eq.~\ref{rtop}. A mystery remains why the $K$ values from the fit are somewhat different from the expected values; still the $K$ values from the fit decrease monotonically with increasing wetness and are predicted if $K_o=0.023$ mm$^2$/min. We show that $dA/dt$ is not constant for any set of $n$-sided bubbles but instead depends on the individual bubble shape and size.

Using these parameters we show how Eq.~\ref{CoarsenEq} also predicts the coarsening behavior of individual bubbles. In particular the coarsening of 6-sided bubbles which is not predicted by von~Neumann's law is determined exclusively by the bubble shape. We show this by solving Eq.~\ref{CoarsenEq} through numerical integration of the circularity data for several 6-sided bubbles from foams of different wetness. The data show the the 6-sided bubbles coarsen with rate changes depending on their shape and this is matched by solutions to the generalized coarsening equation. The shape changes that drive the coarsening rate changes are easily visualized, especially for shrinking 6-sided bubbles, from changes in the film network that cause changes in circularity. Coarsening 6-sided bubbles are the most obvious violations of von~Neumann's law but other $n$-sided also have their coarsening rates reduced due to shape effects. In some cases this leads to obvious non-linear behavior.

We have explored foams here of several wetnesses from relatively dry to relatively wet. Still all the foams are in a quasi-2d dry limit in that the vertices are tri-connected and Plateau's laws hold. It would be interesting to perform experiments where this is no longer the case. We would like to study coarsening of foams with increasing liquid content up to and including the point where bubbles are separated only by liquid faces and there are no thin films. There is a prediction in \cite{SchimmingPRE2017} on how the area of these bubbles should change in time. Coarsening also necessarily relaxes the system and induces rearrangements \cite{GopalDurianPRL2003}. While this study focuses on the dynamics of coarsening, we can also study bubble rearrangements in both dry and wet foams brought on by coarsening.

\begin{acknowledgments}
This work was supported by NASA grant 80NSSC19K0599.
\end{acknowledgments}

\bibliography{HyperRefs}
\end{document}



\title{Supplemental Material for \\ ``Experimentally Testing a Generalized Coarsening Model for Individual Bubbles in Quasi-Two-Dimensional Wet Foams"}


\author{A. T. Chieco \& D. J. Durian}
\affiliation{
   Department of Physics and Astronomy, University of Pennsylvania, Philadelphia, PA 19104-6396, USA 
}

\date{\today}

\begin{abstract}
In this supplement we provide additional information, going beyond the material in the main text, organized as follows.  Section I regards specifications about the sample cell and a schematic cross section of the foam.  Section II regards explicit details about the image analysis methods used for the foam reconstructions.
Section III regards analyses on more bubbles not discussed in the main text.
\end{abstract}



\maketitle




\section{Sample Cell}

\begin{figure*}[!ht]
\captionsetup{justification=raggedright,singlelinecheck=false,margin=20pt,font=small}
\includegraphics[width=5in]{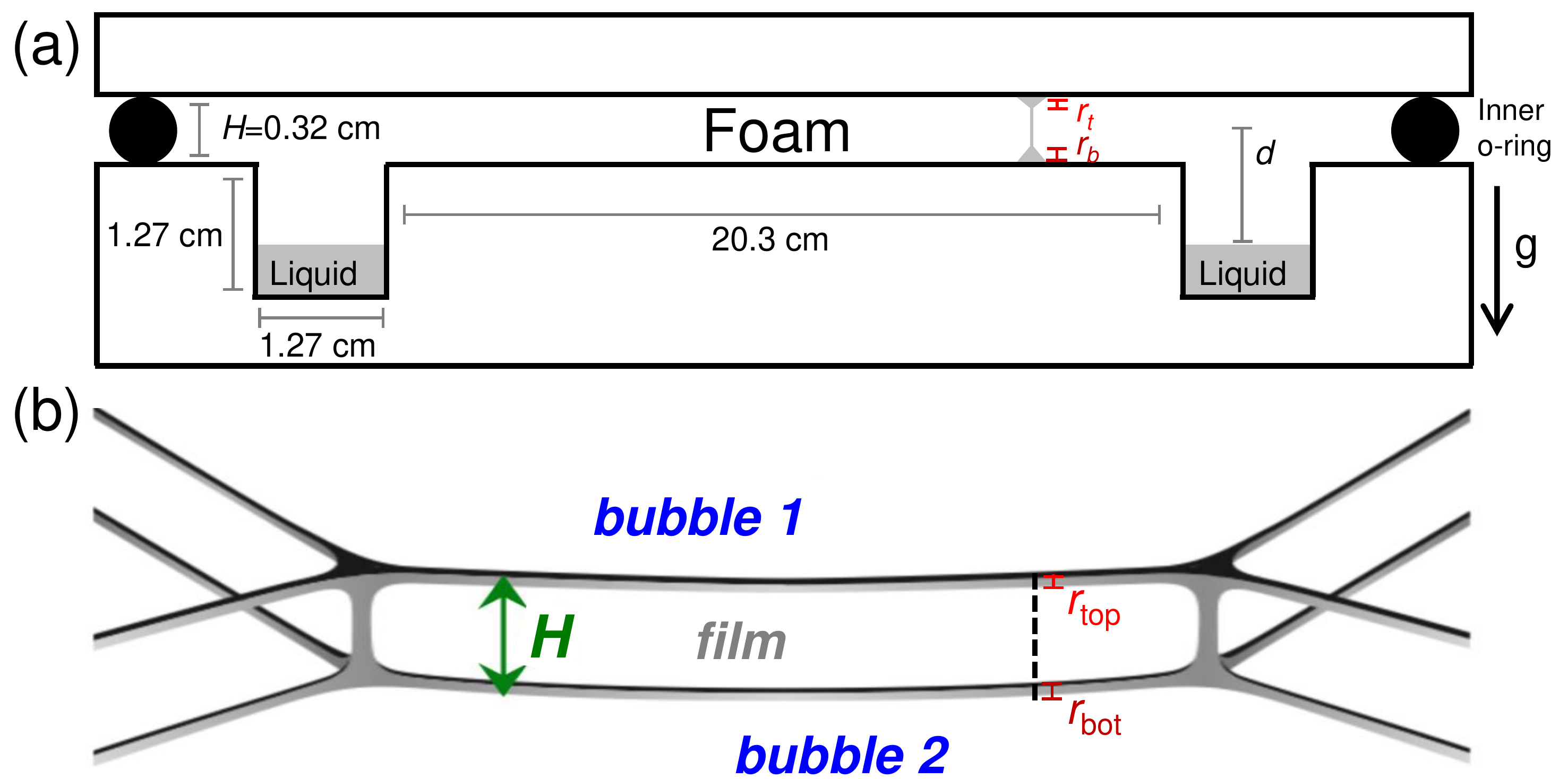}
\caption{Part (a) shows a schematic cross section of the circular constant pressure cell (not to scale). Labeled are the gap between the plates $H$, the distance from the top of the liquid in the reservoir to the middle of the gap between the plates $d$, and the radius of curvature for the Plateau borders along the top and bottom plates, $r_t$ and $r_b$. Part (b) shows a perspective view of two bubbles as they would look inside the sample cell. In this part the thick gray lines are either surface Plateau borders, vertices, or vertical Plateau borders and they represent where the majority of liquid is contained in the foam. Part (a) shows a cross section of the surface Plateau borders and a film corresponding to the black dashed line shown in part (b) if the image in (b) were rotated 90$^o$.  }
\label{foam_cell}
\end{figure*}

Details about our experiment are discussed in Sec.~IIA of the main paper but some information about the sample cell is left out. Here Fig.~\ref{foam_cell}(a) shows a schematic of the sample cell with all measurement specifications necessary to visualize or recreate the sample cell in its entirety. Note that the area of the cell inside the annular trough is over 300 cm$^2$ but the images we analyze are taken from a much smaller central region with dimensions (23.3 $\times$ 15.4)-mm$^2$.

 Fig.~\ref{foam_cell}(a) also shows pictorially how the total amount of foaming solution in the cell affects the wetness of the foam. As depicted the liquid that does not remain in the foam resides in the surrounding liquid reservoir and $d$ is the distance from the top of the liquid in the reservoir to the middle of the gap between the plates. The value of $d$ is inversely proportional to the radius of curvature of the surface Plateau borders $r_t$ or $r_b$ along either the top or bottom plate of the cell, respectively. Also shown is the qualitative size difference between the top and bottom surface Plateau border where $r_t<r_b$ which is apparent from solutions to Eqs.~(4) and (5) in the main paper. 

Fig.~\ref{foam_cell}(b) allows for an intuitive understanding of why wetter foams have slower coarsening rates. It shows a perspective view of the foam as it would look inside the sample cell where both thin films and Plateau borders separate bubbles. Gas diffuses more easily through the thin films than through the surface Plateau borders; wetter foams have Plateau borders that swell with liquid which reduces the area of the thin film and thus the coarsening rate is slower.

\clearpage

\section{Image Analysis}

In this section we explain in more detail the techniques used to reconstruct the foam from the images. We describe the methods we use: to determine the location of the vertices, to find the equations of the circles that represent the films and to ensure the foam reconstructions obey Plateau's laws in subsections A, B, and C, respectively. The last part is especially important for producing very careful reconstructions that accurately reflect the areas of the bubbles in our foams.

\subsection{Locating the Vertices}

\begin{figure*}[!ht]
\captionsetup{justification=raggedright,singlelinecheck=false,margin=20pt,font=small}
\includegraphics[width=3.5in]{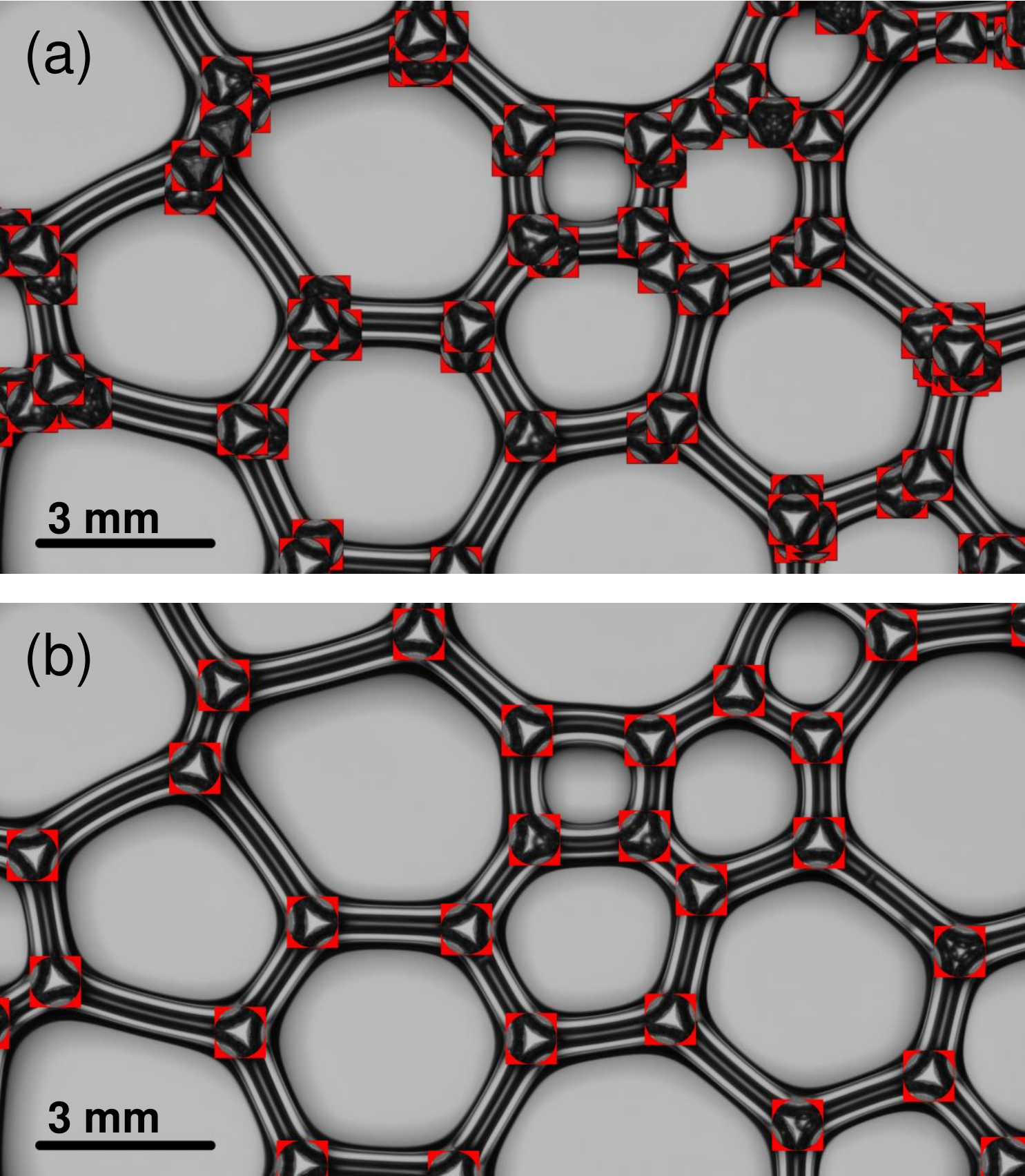}
\caption{Vertex structuring elements at their initial (a) and final (b) positions. The position of the vertex is defined by its $\left(x,y\right)$ coordinate, the orientation $\theta$ of one of its films with the $x$-axis, and the ``type" of vertex defined by features at the center of the structuring element. In part (a) the $\left(x,y\right)$ locations are determined from image processing, and the orientations and types are chose at random. Part (b) shows the vertices at their final positions determined by our Monte Carlo like reconstruction method; there is excellent agreement between the locations and orientations of the structuring elements with the underlying image. }
\label{vertex_overlay}
\end{figure*}

In Sec.~IIB of the main paper we mention that we can not simply binarize, skeletonize and watershed the images of the wet foams in order determine bubble areas like we can for the dry foams. This is because the wet foams have features of varying brightness that make the skeletonized images poor representations of the foam. These features are larger for wetter foams and exist at regions where light channels through the centers of the vertical and surface Plateau borders. Rather than be hampered by these features we take advantage of them to identify the positions of all of the vertices in the foam.

To find the vertex positions we start by binarizing, skeletonizing, and watershedding the images.  The center of the vertex is brighter the surrounding surface Plateau borders and during the image processing these bright features form closed regions in the skeletonized images. Watershedding the skeletonized images results in many regions of various sizes which including the ones at the center of the vertices, the bubbles, and features within the surface Plateau borders. The regions at the center of the vertices are identifiable by their characteristic size and the coordinates of these regions provide an initial guess to the locations of the vertices. All of the initial guesses at vertex locations are shown for one image in Fig.~\ref{vertex_overlay}(a); it is apparent there are far too many locations identified and many of them are incorrect. 

The locations determined from the watershedding basins are close enough to the vertices that they make excellent seeding locations for the structuring elements of a Monte Carlo like reconstruction method. The structuring elements $V$ are $L \times L$ sized sample vertices cut out from the foam images and Fig.~\ref{vertex_overlay}(a) shows examples of these structuring elements at their initial seeding locations. Inspecting the figure shows there are several ``types" of vertices with different features found at their center {\it e.g.} some have a bubble or some are darker. Additionally a circular mask is applied to the structuring elements to make them rotationally symmetric; this is necessary because we rotate the structuring elements and track an angular positions $\theta$ defined by the angle made with the $x$-axis of one of the films that meets at the vertex. Knowing the angle of one film determines the angles of the other two films at the vertex because they are separated by 120$^o$ in accordance with Plateau's laws. We start looking for the location of the vertices with structuring elements whose positions are defined by $P_s=\left[x_s, y_s, \theta, type\right] $ where $x_s$ and $y_s$ are the seeding location determined from watershedding, $0 \leq \theta<120$ is an integer taken at random, and the type is randomly chosen from a user defined list of possible vertex types.

From the initial seeding location the Monte Carlo method samples a new location $P_{\delta}=\left[x_s + \delta_x, y_s + \delta_y, \theta + \delta_{\theta}, type' \right] $ where $\delta_x$, $\delta_y$, and $\delta_{\theta}$ are small perturbations to the locations and angle and $type'$ is an entirely new type of vertex drawn at random. To determine if $P_{\delta}$ is a position that represents a vertex better than $P_s$ we compare the energies of the two locations. The energy of the location is defined by
\begin{equation}
E=\frac{\sum_{ij}{\left(V_{ij}-F_{ij}  \right)^2}}{L^2}.
\label{Energy}
\end{equation}
Here $F_{ij}$ is an $L \times L$ cut out of the foam image with a circular mask applied that corresponds to the location of the structuring element and the values $i$ and $j$ run over corresponding pairs of pixels between $V$ and $F$. If $E_{\delta}<E_s$ then the vertex position updates to the new coordinates otherwise it remains at the old ones. This probing and updating is repeated for five thousand steps. At the end of the interrogation any structuring element whose $E$ value is higher than some threshold is removed as are duplicates. A video accompanying the supplemental material shows the entire sampling and removal process. Fig.~\ref{vertex_overlay}(b) shows the structuring elements at their final location, orientation, and type and there is remarkable agreement with the underlying image.

\subsection{Finding the Films}

\begin{figure*}[!ht]
\captionsetup{justification=raggedright,singlelinecheck=false,margin=20pt,font=small}
\includegraphics[width=3.5in]{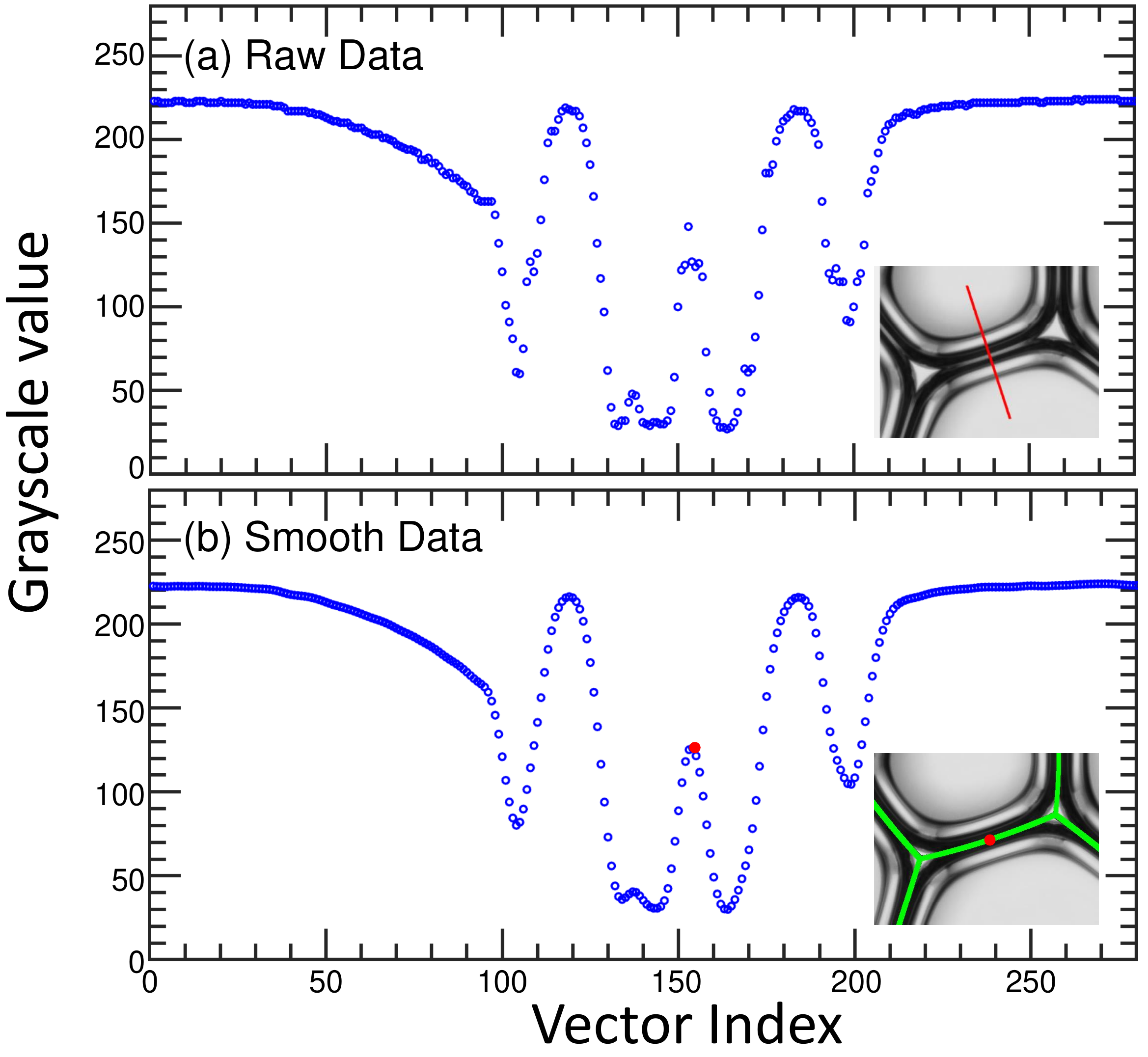}
\caption{Grayscale values found from a vector crossing over a surface Plateau border where the data is either unfiltered (a) or smoothed over a Gaussian window (b). In part (a) the inset shows the pixel locations that correspond to the data and the vector index is 0 at the top left of the line. Part (b) show the data smoothed over a Gaussian window and the red dot in the plot is the vector index that corresponds to center of the film. The inset of part (b) shows the location in the image determined to be the center of the film and the green lines are the reconstructions of the film network.     }
\label{films_find}
\end{figure*}

According to Plateau's laws the films connecting the vertices in quasi 2-dimensional foam are arcs of circles. The center and radius $(x_c, y_c, \mathcal{R})$ of the circles that connect any pair of vertices are defined by the vertex locations and a point midway between the two vertices in the center of the film. To determine this third point we draw a line that perpendicularly bisects the line that connects the two vertices. This perpendicular bisector has length $L$ chosen to be long enough to pass through the entire film connecting the vertices but is not so long that is encounters other films; the line is drawn $L/2$ in either direction away from the midpoint between the vertices. The points on this line are converted into vector indices and we find the grayscale values for the image at these locations. An example of found grayscale values is plotted in the main part of Fig.~\ref{films_find}(a) and the inset shows the corresponding line.

For all films the features most important in determining the center of the film are the first local minima found as we raster across vector indices from low to high and from high to low. These minimia ensure that we are probing for a point inside the surface Plateau border. As we continue moving from low to high and high to low there are two local maxima that further reduce the potential candidates for the point at the center of the film. Once our search is reduced to the indices between these two peaks we look for the vector index of the largest local maximum and this is the center of the film. Because we are looking for local minima and maxima it is easiest to analyze data smoothed through a Gaussian window; the smoothing windows has a width about 10 pixels which is same size as the bright band at the center of the surface Plateau border. The smoothed data is plotted in Fig.~\ref{films_find}(b) and the vector index corresponding to the center of the film is highlighted in red. The inset in Fig.~\ref{films_find}(b) shows where the the index is on the image as a red circle and the underlying green lines are the reconstructed film network.

\subsection{Making Reconstructions Agree with Plateau's Laws }

\begin{figure*}[!ht]
\captionsetup{justification=raggedright,singlelinecheck=false,margin=20pt,font=small}
\includegraphics[width=3.5in]{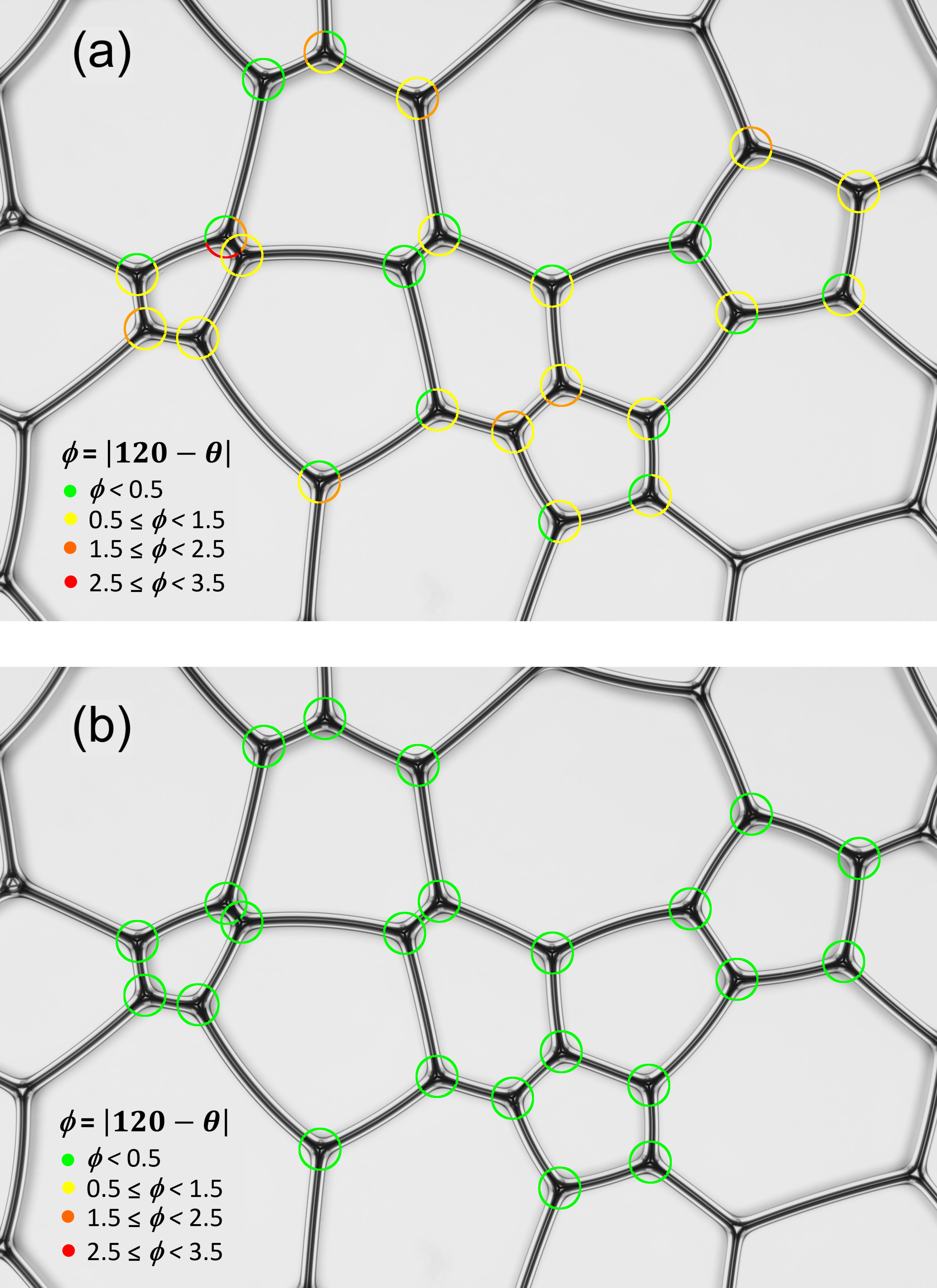}
\caption{Part (a) and (b) show a range in the excess angle $\phi$ between neighboring tangent lines represented by the color of the circular arcs that connect them. These lines are tangent to the reconstructed films that meet at a vertex and according to Plateau's laws should all be separated by 120$^o$. Part (a) shows the angle between tangent lines after the film network is identified for the first time; there is a range of $\phi$ values and some are far from 120$^o$. Part (b) shows the angle between tangent lines after the vertices are shifted to better satisfy Plateau's laws; all of the excess angles are shown to be within half a degree of 120$^o$ but in fact they are much closer as evidenced in Fig.~\ref{phi_distributions}.  }
\label{phis_foams_pic}
\end{figure*}

One of Plateau's laws already discussed is films meet at a vertex and 120$^o$ which we use to find the vertex locations and identify neighbors. 
What actually meets at 120$^o$ are the lines tangent to the circular arcs that meet at the vertices so to find these angles we need to reconstruct the film network first. Once the film network is reconstructed the tangent lines at the vertices are identified and the angles between neighboring tangent lines are determined. The reconstructions match the foam well by eye but if we evaluate an ``excess angle" $\phi=| 120-\theta |$ we find that some of the vertices are actually far from mechanical equilibrium. This is shown in Fig.~\ref{phis_foams_pic}(a) where arcs connecting neighboring tangent lines are colored based on the excess angle between those tangent lines. 

The reason vertices have films that meet at 120$^o$ is because they are in mechanical equilibrium; if the tangents do not meet at 120$^o$  there would be  some net force at the vertex. We treat the reconstructions as if this were the case and the excess force is determined as
\begin{eqnarray}
	F_x &=& \sum_{i}{\text{cos} \left( \psi_i \right)} \\
	F_y &=& \sum_{i}{\text{sin} \left( \psi_i \right)}
\end{eqnarray}
where $F_x$ is a force in the $x$-direction, $F_y$ is the force in the $y$-direction and $\psi$ is the angle made with the $x$-axis by the tangents at the vertex if the vertex were shifted to the point $\left(0,0\right)$. These forces are multiplied by some small constant scaling coefficient to convert them into a displacement for the vertex. After the vertex is shifted we redo the process of finding the films, calculating the tangent lines, and checking the excess angle for the new vertex locations. This is repeated until $\phi <0.01^o$ for all of the excess angles. A video accompanying the supplemental material shows this relaxation to mechanical equilibrium which is usually accomplished in about 20 steps. We show the final results of the tangent lines in the foam in Fig.~\ref{phis_foams_pic}(b) where it is clear all the tangents have converged to angle close to 120$^o$.

\begin{figure*}[!ht]
\captionsetup{justification=raggedright,singlelinecheck=false,margin=20pt,font=small}
\includegraphics[width=3.5in]{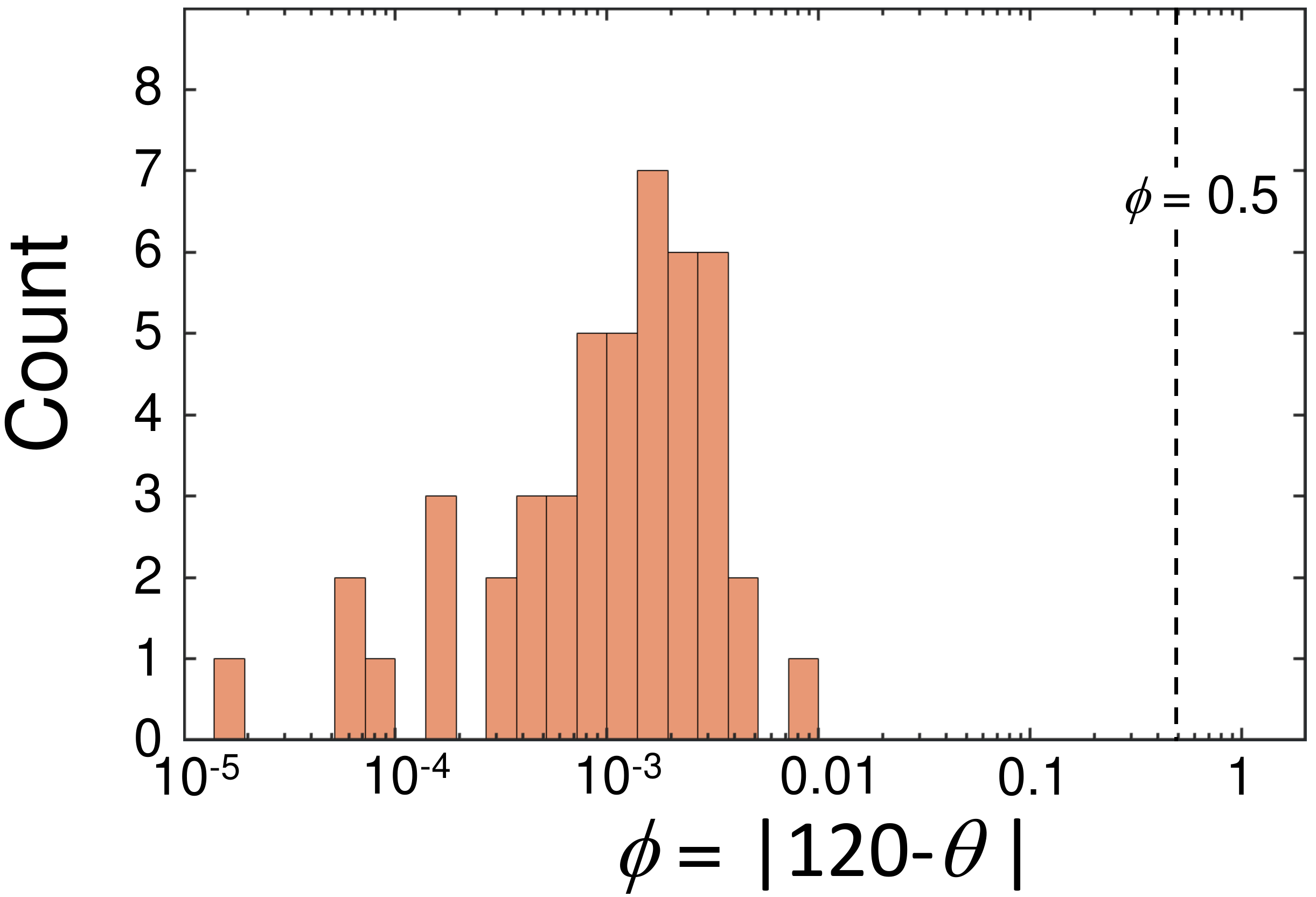}
\caption{The distribution of the excess angle $\phi$ after the vertices have been relaxed to better match Plateau's laws. $\phi=0.5^o$ is the value used in Fig.~\ref{phis_foams_pic}(b) to label the excess angles and it is marked here with a dashed line; in actuality the $\phi$ values achieved by our relaxation technique are much smaller and all satisfy the condition $\phi<0.01^o$. }
\label{phi_distributions}
\end{figure*}

Limitations exist on representing the actual excess angles on the image so we also plot the distribution of excess angles in Fig.~\ref{phi_distributions}. Here it is clear the $\phi$ are not only  smaller than the $\phi<0.5$ as shown in the Fig.~\ref{phis_foams_pic}(b) but in fact many of the $\phi$ values are even orders of magnitude smaller than the convergence criteria of $\phi<0.01$. Getting the reconstructions to match with Plateau's laws was an important and necessary step in getting the correct areas and coarsening rates for individual bubbles.

\clearpage

\section{Additional Data}

In this section we show additional data not shown in the main paper. Sec.~IIIA  Fig.~\ref{A_and_C_six} and Fig.~\ref{A_and_C_many} demonstrate how the circularity of a bubble influences its coarsening for many 6-sided and many $n$-sided bubbles, respectively. Both figures are made from foam with $d=5.9$mm and the data for $n=6$ in Fig.~\ref{A_and_C_many} is the same data from Fig.~\ref{A_and_C_six}. In Sec.~IIIB we present plots similar to the main paper Fig.~6 but for 5-sided and 7-sided bubbles in Fig.~\ref{5_sided_coarsening} and Fig.~\ref{7_sided_coarsening}, respectively.

\subsection{Bubble Areas and Circularities for Many Bubbles}

\begin{figure*}[!ht]
\captionsetup{justification=raggedright,singlelinecheck=false,margin=20pt,font=small}
\includegraphics[width=3.5in]{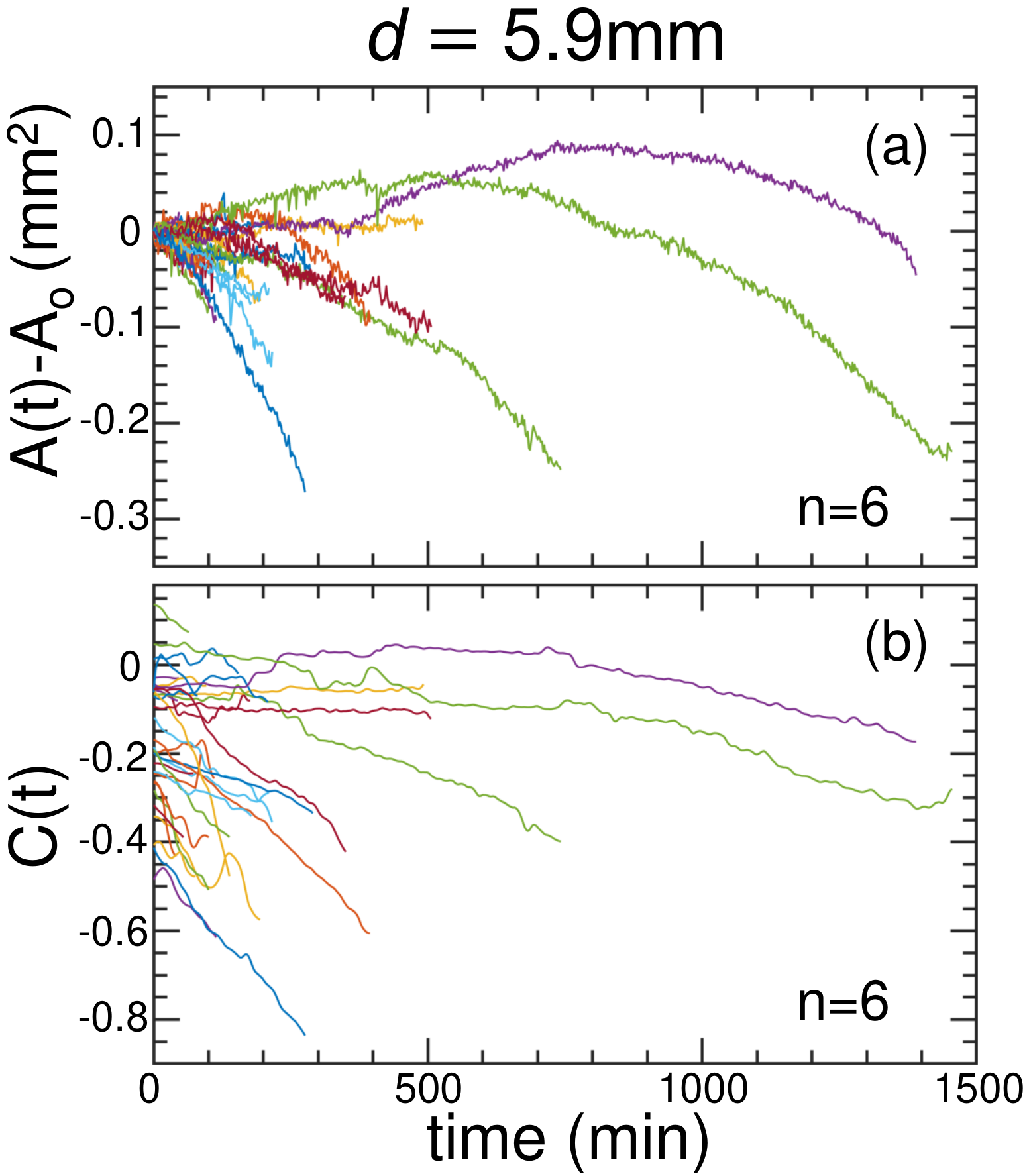}
\caption{Change in area (a) and circularity (b) versus time for many 6-sided bubbles from a foam with wetness as labeled. Curves that are the same color in part (a) and part (b) are data for the same bubble. 6-sided bubbles with $C\left( t \right)<0$ shrink, $C\left( t \right)>0$ grow and some bubbles coarsen non-monotonically depending on the sign of $C\left( t \right)$.}
\label{A_and_C_six}
\end{figure*}

In the main paper we showed for a few bubbles how their circularity specifically determined their coarsening behavior. This was demonstrated through solutions to the generalized coarsening equation using the bubble specific circularity data. We show here for many bubbles that circularity is causing deviations from von~Neumann like behavior by either driving coarsening in 6-sided bubbles or causing deviations from linear behavior for other $n$-sided bubbles. For 6-sided bubbles we plot the change in area and circularity versus time in Fig.~\ref{A_and_C_six}(a) and (b). Because these bubbles have $dA/dt = 6KnrC\left(t\right) / \sqrt{3 \pi A\left(t\right)}$ only the circularity can drive coarsening; the change in area and the sign and magnitude of the circularity  are shown to correspond because curves with the same color in part (a) and (b) are data for the same bubble.  6-sided bubbles with negative circularity shrink, positive circularity grow and some bubbles behave non-monotonically depending on the sign of $C\left( t \right)$. 

\begin{figure*}[!ht]
\captionsetup{justification=raggedright,singlelinecheck=false,margin=20pt,font=small}
\includegraphics[width=3.5in]{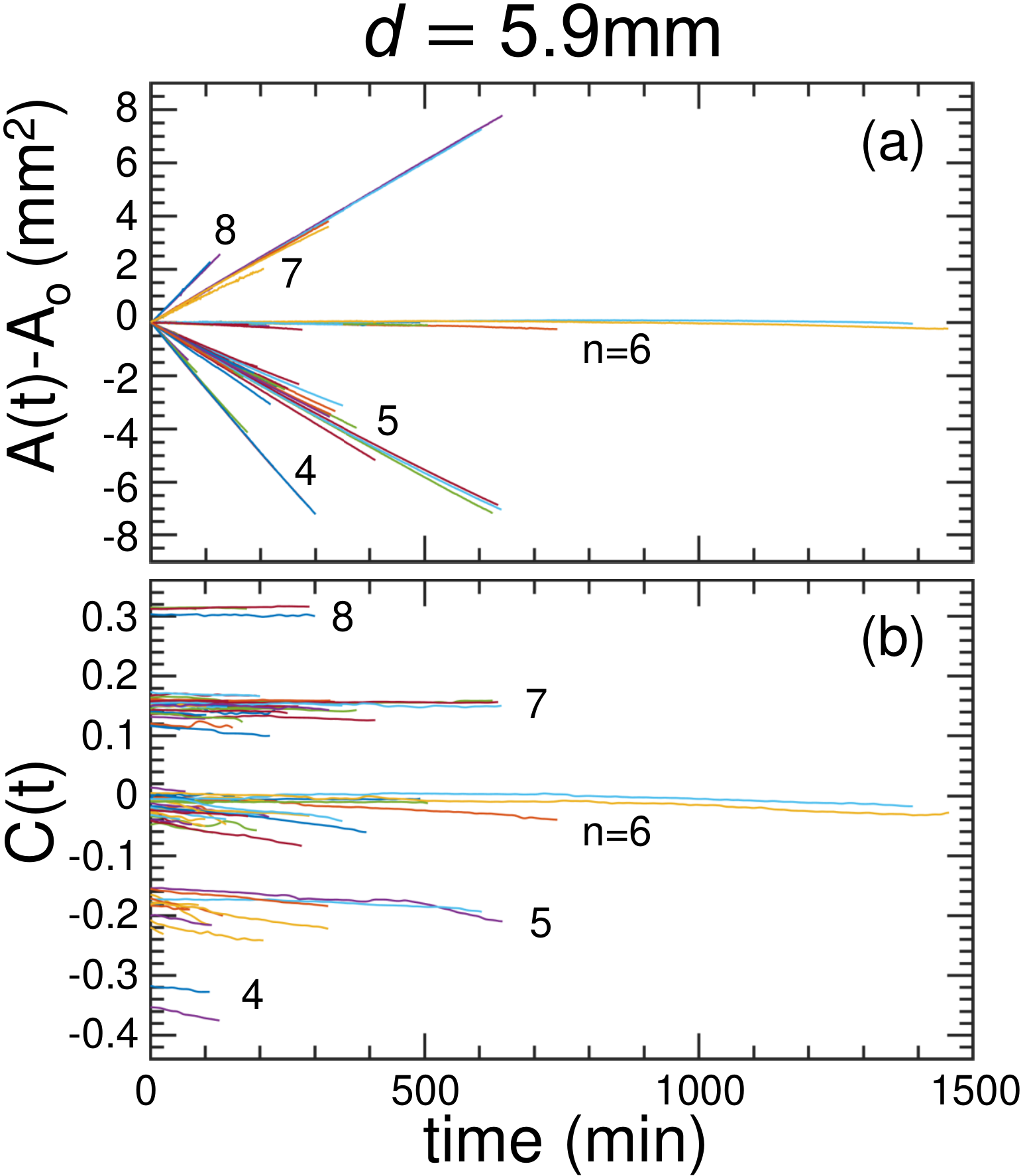}
\caption{Change in area (a) and circularity (b) versus time for many $n$-sided bubbles from a foam with wetness as labeled. Curves that are the same color in part (a) and part (b) are data for the same bubble. Part (a) shows $n$-sided bubbles with different circularities will coarsen at different rates which is in accordance with the generalized coarsening equation but violates von~Neumann's law. Part (b) shows that bubbles with the same number of sides have circularities that are relatively quantized }
\label{A_and_C_many}
\end{figure*}

For many $n$-sided bubbles we plot the change in area and circularity versus time in Fig.~\ref{A_and_C_many}(a) and (b). In this figure it is more difficult to tell exactly how much the circularity affects the coarsening because the effect is small compared to the overall area change. However the data show how the coarsening rates are different for bubbles with the same number of sides depending on the bubble shape; this is especially apparent for the 5-sided bubbles where the area data fan out depending on how much the circularity slows the coarsening.

\clearpage

\subsection{Individual Bubble Coarsening}

\begin{figure*}[!ht]
\captionsetup{justification=raggedright,singlelinecheck=false,margin=20pt,font=small}
\includegraphics[width=6in]{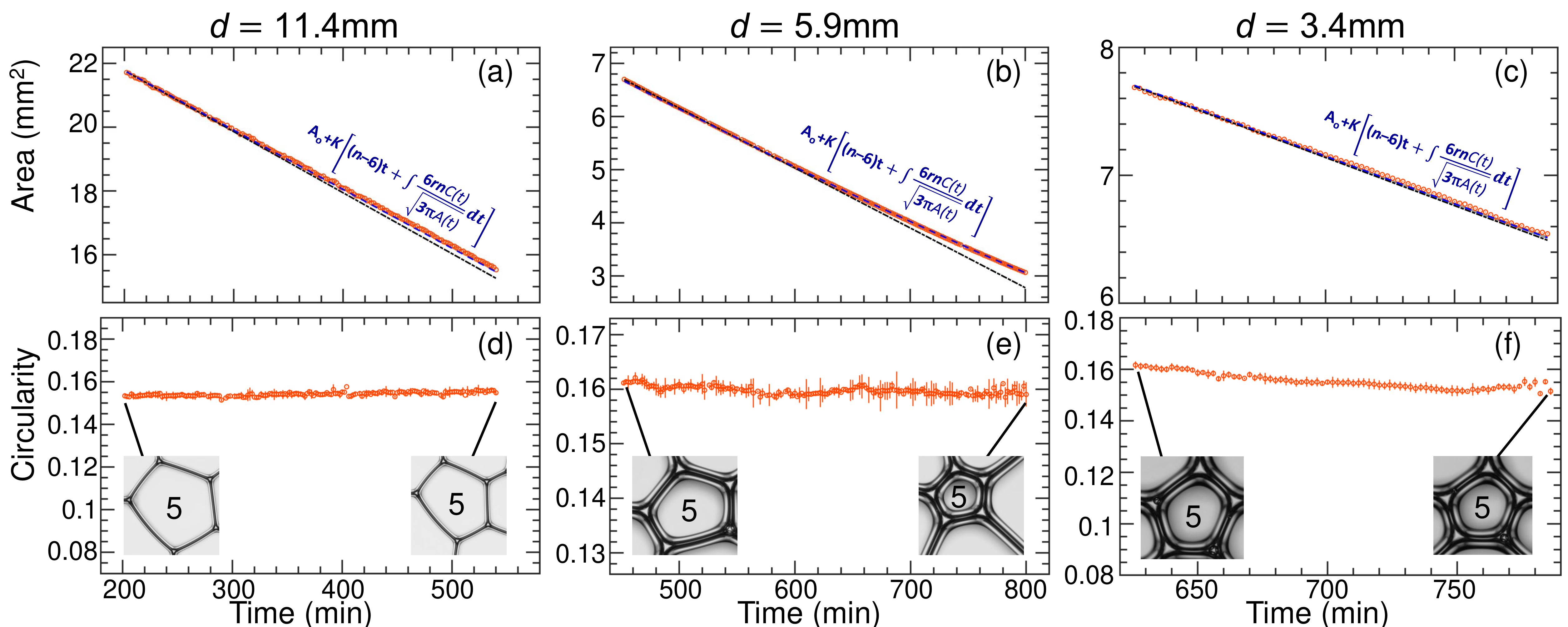}
\caption{The area (a-c) and circularity (d-f) versus time for individual 5-sided bubbles. The bubbles come from foams of increasing wetness from left to right as labeled by decreasing $d$. The images in parts (d-f) show the bubble at the times pointed to by the black lines. The von~Neumann expectation is that 5-sided bubbles shrink linearly in time but this is not the case for the bubbles highlighted here; in parts (a-c) the black dash-dotted lines are tangent to the data at early times but the bubble coarsening is non-linear and the data deviate from the line at late times. This behavior is due to the bubble circularity where 5-sided bubbles shrink more slowly because $C>0$. The blue dashed lines are the numerically integrated solutions to the generalized coarsening equation using the bubble circularity data as well as values of $K$ and $r$ corresponding to the bubble wetness. The gray swaths are generated similarly, incorporating statistical uncertainty in $C$ as well as uncertainty in the values of $K$ and $r$.}
\label{5_sided_coarsening}
\end{figure*}

\begin{figure*}[!ht]
\captionsetup{justification=raggedright,singlelinecheck=false,margin=20pt,font=small}
\includegraphics[width=6in]{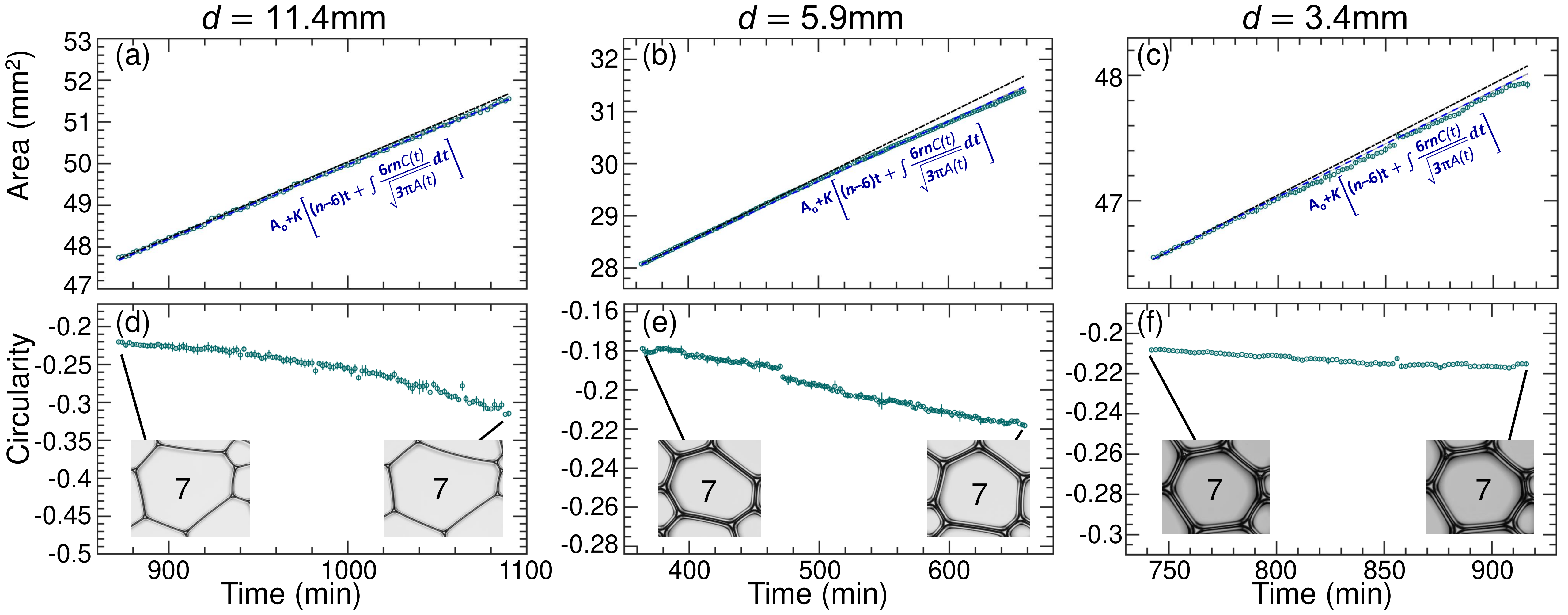}
\caption{The area (a-c) and circularity (d-f) versus time for individual 7-sided bubbles. The bubbles come from foams of increasing wetness from left to right as labeled by decreasing $d$. The images in parts (d-f) show the bubble at the times pointed to by the black lines. The von~Neumann expectation is that 7-sided bubbles grow linearly in time but this is not the case for the bubbles highlighted here; in parts (a-c) the black dash-dotted lines are tangent to the data at early times but the bubble coarsening is non-linear and the data deviate from the line at late times. This behavior is due to the bubble circularity where 7-sided bubbles grow more slowly because $C<0$. The blue dashed lines are the numerically integrated solutions to the generalized coarsening equation using the bubble circularity data as well as values of $K$ and $r$ corresponding to the bubble wetness. The gray swaths are generated similarly, incorporating statistical uncertainty in $C$ as well as uncertainty in the values of $K$ and $r$.}
\label{7_sided_coarsening}
\end{figure*}

Additional demonstrations of the quantitative validity of Eq.~(2) are shown here with solutions for 5-sided and 7-sided bubbles from foams of various wetness.. 
We again numerically integrate the generalized coarsening using the displayed $C(t)$ data in order to obtain $A(t)$ versus $t$ along with the fitted values of $K$ and $r$ discussed in Sec.~III of the main paper. The resulting predictions for $A(t)$ versus $t$ are displayed as dashed curves with a surrounding gray swath that reflects statistical uncertainties in $K$, $r$, and $C$. Data for 5-sided and 7-sided bubbles are plotted in Fig.~\ref{5_sided_coarsening} and Fig.~\ref{7_sided_coarsening}, respectively.

